\documentclass[aps,prb,twocolumn,superscriptaddress,showpacs,amsmath,amssymb,longbibliography]{revtex4-1}
\usepackage[english]{babel}
\usepackage{amsmath}
\usepackage{bm}
\usepackage{graphicx,bbm}
\usepackage{times}
\usepackage{epsfig,psfrag} 
\begin{document} 
\title{Self-consistent approach to many-body localization and subdiffusion}
\author{P. Prelov\v{s}ek}
\affiliation{Jo\v zef Stefan Institute, SI-1000 Ljubljana, Slovenia}
\affiliation{Faculty of Mathematics and Physics, University of Ljubljana, SI-1000 
Ljubljana, Slovenia}
\author{J. Herbrych}
\affiliation{Department of Physics and Astronomy, The University of Tennessee, 
Knoxville, Tennessee 37996, USA}
\affiliation{Materials Science and Technology Division, Oak Ridge National Laboratory, 
Oak Ridge, Tennessee 37831, USA}

\date{\today}
\begin{abstract}
An analytical theory, based on the perturbative treatment of the disorder and extended 
into a self-consistent set of equations for the dynamical density correlations, is 
developed and applied to the prototype one-dimensional model of many-body localization. 
Results show a qualitative agreement with numerically obtained dynamical structure 
factor in the whole range of frequencies and wavevectors, as well as 
across the transition to the nonergodic behavior. The theory reveals the singular nature of the 
one-dimensional problem, whereby on the ergodic side the dynamics is subdiffusive with 
dynamical conductivity $\sigma(\omega) \propto |\omega|^\alpha$, i.e., 
with vanishing d.c. limit $\sigma_0=0$ and $\alpha<1$ varying with disorder,
while we get $\alpha >1$ in the localized phase.

\end{abstract}

\maketitle

\section{Introduction}
 
Many-body localization (MBL) is a challenging phenomenon involving the interplay of 
disorder and particle interaction (correlations). In the fermionic systems it has been 
proposed as an extension of the single-particle Anderson localization 
\cite{anderson58,mott68}, remaining qualitatively valid at finite interactions 
\cite{fleishman80,basko06} and at large enough disorder even at high temperature $T$~ 
\cite{oganesyan07}. In contrast to normal (ergodic) systems, the MBL state should reveal 
vanishing d.c. transport 
\cite{berkelbach10,barisic10,agarwal15,gopal15,lev15,steinigeweg15,barisic16} as well as a 
nonergodic time evolution of correlation functions and of quenched initial states 
\cite{pal10,serbyn131,deluca13,huse14,vosk13,vosk14,serbyn14,vasseur15}. The vanishing of d.c. 
mobility \cite{kondov15} and the nonergodic decay of the initial density profile 
\cite{schreiber15,bordia16,luschen16} have been the main experimental signatures of the 
MBL in fermionic cold-atom systems. 

The dynamical structure factor $S(q,\omega)$ is the obvious observable to characterize 
the one-dimensional (1D) system undergoing the ergodic-nonergodic (MBL) transition. 
Theoretical studies so far concentrated mostly on the uniform (wavevector $q\to 0$) 
response as, e.g., contained in the optical conductivity $\sigma(\omega)$ and its d.c. 
limit $\sigma_0$ 
\cite{berkelbach10,barisic10,agarwal15,gopal15,lev15,steinigeweg15,barisic16}. In this 
connection, a challenging question is the possibility of subdiffusive dynamics, 
\cite{agarwal15,gopal15,luitz16,znidaric16,luitz116,luschen16,vosk15} which implies vanishing 
d.c. transport, e.g., $\sigma_0 = 0$ but anomalous low-$\omega$ dependence of the optical 
conductivity $\sigma \propto |\omega|^\alpha$ with $\alpha <1$. On the other hand, in 
the cold-atom experiments so far more accessible are density correlations with $q=\pi$ 
\cite{luitz15,schreiber15,mierzejewski16}, as measured via the time-dependent imbalance 
\cite{schreiber15,bordia16,luschen16}.

In this paper we first present results for $S(q,\omega)$ within the prototype disordered 1D
model of interacting spinless fermions, displaying
the whole range of wavectors $q = [0,\pi]$, 
as obtained with a numerical calculation at $T \to \infty$ on small finite-size systems with up 
to $L =24$ sites. We show that it is convenient and informative to analyse the $S(q,\omega)$ 
spectra in terms of memory functions, representing the corresponding dynamical
conductivity $\sigma(q,\omega)$ and even further the current decay-rate function
$\Gamma(q,\omega)$. Such quantities reveal more clearly the transition to the MBL
regime, as well as the behavior in the case of subdiffusion. 

We further introduce for the same model an analytical theory, based on the perturbative treatment 
of the current-decay function $\Gamma(q,\omega)$ and extended to a self-consistent (SC) evaluation of 
density-correlation function $\phi(q,\omega)$. The theory reveals the specific nature of the 1D problem, 
leading to a singular coupling 
between $q\to 0$ density and energy diffusion modes. Still, the solution of the SC 
equations with an additional cut-off simulating a finite system size $L^*$ shows 
qualitative (and at weaker disorder even quantitative) agreement with numerically 
obtained results for $S(q,\omega)$ and related $\sigma(q,\omega)$. 
Moreover, the finite-size scaling of SC results reveals in the 
ergodic phase the subdiffusive dynamics, consistent with 
$\sigma(\omega \to 0) \sim |\omega|^\alpha$ with $\alpha <1$. The MBL transition at 
critical disorder $W=W_c$ is thus determined by a dynamical exponent $\alpha=1$, while
the MBL phase is characterized by $\alpha>1$ and a finite dielectric polarizability 
$\chi_{\mathrm{d}}$ of the insulating system. 

The paper is organized as follows: In Sec.~II we present the model and the general formalism
for density dynamical susceptibility $\chi(q,\omega)$, which is related to
generalized dynamical conductivity $\sigma(q,\omega)$ and further to the current 
decay-rate function $\Gamma(q,\omega)$. In Sec.~III we present results for $S(q,\omega)$, 
obtained via numerical exact-diagonalization technique for $T \to \infty$ on finite chains 
for all available $q$ . Results of $\sigma(q,\omega)$ and $\Gamma(q,\omega)$, obtained 
 with help of formalism introduced in Sec.~II, are also presented.
This allow for connection with previous studies of, e.g., optical
conductivity $\sigma(\omega)$ and also a motivation as well as a stringent test for the 
proposed analytical theory. In Sec.~IV we introduce analytical approximations, 
based on the perturbative treatment of $\Gamma(q,\omega)$. 
Furthermore, with some additional simplifications, solution for $\Gamma(q,\omega)$ is 
extended into a SC set of equations. 
Numerical results of these equations are presented and 
commented in Sec.~V. Besides the qualitative agreement with finite-size results, we put the
emphasis on the low-$\omega$ regime where the SC equations appear to be singular in 1D. 
Scaling an effective chain length $L^*$ we show that in the ergodic regime the solutions are
consistent with an interpretation in terms of a subdiffusion phenomenon. Conclusions, critical
reflections on the method and results are given in Sec.~VI. 

\section{Dynamical density correlations}
 
We consider the prototype (standard) model of MBL, the 1D system of interacting spinless 
fermions with random local potentials, 
\begin{equation}
H = - t \sum_{i} \left( c^\dagger_{i+1} c_i + \mathrm{H.c.}\right)
+ V \sum_i n_{i+1} n_{i} + \sum_i \epsilon_i n_i\,.
\label{tvw}
\end{equation}
As usual, we assume quenched disorder with the uniform distribution 
$-W < \epsilon_i <W$ and in the numerical analysis the system at half-filling, i.e., 
$\bar n=1/2$. We further-on use $t=1$ as the unit of energy. While most numerical 
results so far are for $V=2$ (corresponding to isotropic Heisenberg model 
\cite{berkelbach10,agarwal15,gopal15,lev15,steinigeweg15,barisic16}), we use for 
the demonstration $V=1$ enabling closer comparison with the analytical theory. Since $T$ 
should not play an essential role in the MBL problem, studies are adapted to the limit 
$\beta =1/T \to 0$, which simplifies the analytical as well as the numerical approach.

Our analysis deals with dynamics of the density operator 
\begin{equation}
n_q = \frac{1}{\sqrt{L}} \sum_i \mathrm{e}^{\imath qi} n_i\,,
\end{equation}
 at arbitrary wavevector $q$, as defined by 
the dynamical susceptibility $\chi(q,\omega)$, and related relaxation function
$\phi(q,\omega)$, 
\begin{eqnarray}
\chi(q,\omega)&=& \imath \int_0^\infty\mathrm{d} \tau\,
\mathrm{e}^{\imath\omega \tau} \langle [n_q(\tau),n_{-q}] \rangle\,, \nonumber \\
\phi(q,\omega)&=& \frac{1}{\omega} [\chi(q,\omega)-\chi^0(q)]\,,
\label{chiqw}
\end{eqnarray}
with its static (thermodynamic) value $\chi^0(q)$ (see formal background and definitions 
in Appendix A), which in normal ergodic systems satisfies $\chi^0(q) = \chi(q,\omega \to 0)$. 

In a homogeneous system $\langle ..\rangle$ denotes the canonical thermodynamical 
average. In a disordered system we perform in addition the averaging over all random 
configurations of $\epsilon_i$. We have to stress that $n_q$ is a macroscopic operator 
(not a local one), and in the following analysis we study only such quantities. 
This implies that dynamical correlations functions, as defined by Eq.~\eqref{chiqw}, are 
expected to be self-averaging, i.e., the configuration averaging is in principle not required in the macroscopic 
limit of $L \to \infty$. Here, we rely on a similarity with the treatment of Anderson localization
of NI electrons \cite{vollhardt80,vollhardt801} as well on recent analysis of 
sample-to-sample fluctuations of $\sigma(\omega)$ in the same MBL model \cite{barisic16}. Nevertheless, this
aspect has still to be critically examined when taking the limit $\omega \to 0$ since the 
fluctuations at larger disorder can be singular (referred in 1D as the Griffiths effect 
of rare but large random deviations \cite{agarwal15,gopal15,gopal16}). In particular, this is
the relevant question within the nonergodic (MBL) phase.

The advantage of above formulations is that it remains meaningful even in 
nonergodic cases where $\chi^0(q) > \chi(q,\omega \to 0)$ 
\cite{pirc74,gotze79,vollhardt80,vollhardt801}, as 
expected within the MBL regime. It is helpful to represent and analyse $\phi(q,\omega)$ 
in terms of complex memory functions \cite{forster95} (see formal derivation and relations 
in Appendix B),
\begin{eqnarray}
\phi(q,\omega)&=&\frac{- \chi^0(q)}{\omega+M(q,\omega)}, \nonumber \\
M(q,\omega) &=& \imath \frac{g_q^2}{ \chi^0(q) } \sigma (q,\omega)\,,
\label{mem}
\end{eqnarray}
related to the $q$-dependent conductivity $\sigma(q,\omega)$ via the continuity equation 
$[H,n_q] = g_q j_q$, where $g_q = 2 \sin(q/2)$ and $j_q$ is the current operator for 
given $q$. It should be noted that $\sigma(q,\omega)$ has the usual meaning only in the limit 
$q \to 0$, where $\sigma(\omega) = \mathrm{Re}\,\sigma(q \to 0,\omega)$ is the optical 
conductivity. We make a further step and define the current relaxation-rate function 
$\Gamma(q,\omega)$ as 
\begin{equation}
\sigma (q,\omega) = \frac{\imath\chi^0_{j}(q)}{\omega+\Gamma(q,\omega)}\,,
\label{sigqw}
\end{equation}
where $\chi^0_j(q)$ is the static current susceptibility. We note that 
$\gamma(q,\omega) =\mathrm{Im}\,\Gamma(q,\omega)$ is (at $\beta \to 0$) independent of 
$\beta$ and has the meaning of the effective current relaxation rate at $\omega \to 0$. 

The limit $\beta \to 0$ allows also for analytical evaluation of static quantities, in 
particular,
\begin{eqnarray}
\chi^0(q) &=& \beta \bar n(1-\bar n) =\chi^0\,, \nonumber \\
\chi_{j}^0(q) &=& \beta 2 t^2 \bar n(1-\bar n) = \chi^0_j\,.
\label{chiq0}
\end{eqnarray}
With known static quantities, Eq.~\eqref{chiq0}, the relation between $\phi(q,\omega)$ and,
e.g., $\Gamma(q,\omega)$ is thus unique and exact (not depending on approximations
introduced later-on), and can be used in any direction
provided that one of both quantities is evaluated. It is worth mentioning that 
Eq.~\eqref{mem} together with Eq.~\eqref{sigqw} resemble continued fraction expansion of 
frequency moments of complex correlation functions. Such a approach was recently 
used in Ref.~\onlinecite{khait16} to numerically evaluate 
optical conductivity $\sigma(\omega)$ in strong disorder limit. Here, we develop
analytical theory for first three moments of such a series (see Section~IV). 

\section{Numerical finite-size results}

\begin{figure}[!htb]
\includegraphics[width=1.0\columnwidth]{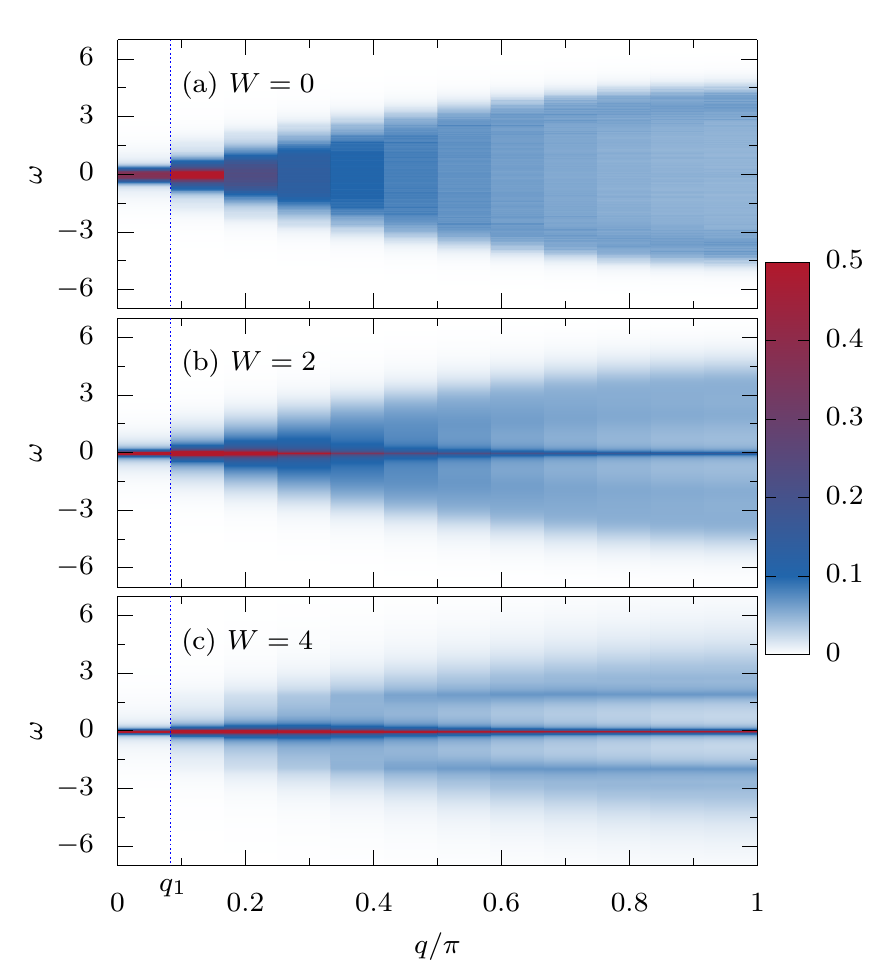}
\caption{(Color online) High-temperature $\beta\to0$ dynamical structure factor 
$S(q,\omega)$, as calculated with MCLM for 
$L=24$, $V=1$, (a) $W=0$, (b) $W=2$, and (c) $W=4$ for all $q=[0,\pi]$.}
\label{figS1}
\end{figure}

We note the relation of above quantities to standard dynamical structure 
factor $S(q,\omega)$, which is at $\beta\to 0$ given by 
\begin{equation}
 \mathrm{Im}\,\phi(q,\omega) = \pi\beta S(q,\omega)\, .
\end{equation}
Before introducing the analytical method, we comment on numerical finite-size results, 
which serve later-on as a test for the proposed analytical theory. 
The dynamical quantity calculated directly is 
$S(q,\omega)$, whereby we employ
the microcanonical Lanczos method (MCLM) on finite systems at $\beta \to 0$ 
\cite{long03,prelovsek13}. In Fig.~\ref{figS1} we present 
characteristic results for $S(q,\omega)$ for $L=24$, $V=1$, $W=0,2,4$ in the whole 
range of wavevectors $q=[0,\pi]$. They already allow for some rough distinction
of dynamical density correlations in three regimes: (a) At $W = 0$ $S(q,\omega)$
is the response of the homogeneous 1D chain of interacting spinless fermions.
Due to integrability of such a model, even at $\beta \to 0$ the response has close 
analogy to non-interacting (NI) fermions (i.e., at $V=0$) \cite{herbrych12}. In particular, the 
$S(q,\omega)$ has no diffusion pole and is quite featureless (at $\beta \to 0$)
in the interval $\omega < 4 \sin(q/2)$. (b) At weak disorder $ 0<W = 2 <W_c \sim 3.5$ 
~\cite{lev15} an additional
feature is a diffusion (or diffusion-like, as discussed later-on in relation to subdiffusion) 
pole which has a finite width $\delta \omega \propto q^2$ and is well visible 
at small $q \ll \pi/2$. (c) For large disorder $W = 4> W_c$ the response becomes
singular at all $q$ and $S(q,\omega \sim 0) = S_q \delta(\omega)$ 
shows a finite stiffness $S_q >0$, being a hallmark of MBL regime. 

Since $S(q,\omega)$ is quite singular function (at least for 
$q \to 0$), it is helpful to extract the corresponding $\sigma(q,\omega)$ and 
$\Gamma(q,\omega)$ via Eqs.~\eqref{mem}, \eqref{sigqw}. To this purpose we first 
calculate complex $\phi(q,\omega)$ from $S(q,\omega)$. Next, with known 
$\chi^0, \chi^0_j$ using Eqs.~\eqref{mem},\eqref{sigqw} we evaluate $\sigma(q,\omega)$ 
and $\Gamma(q,\omega)$. In the numerical procedure it is crucial to have high frequency $\omega$ 
resolution of MCLM results, which are obtained by employing $N_L \sim 10^4$ Lanczos 
steps in order to get $\delta \omega \lesssim 0.003$ of $S(q,\omega)$ spectra. 

\begin{figure}[!htb]
\includegraphics[width=1.0\columnwidth]{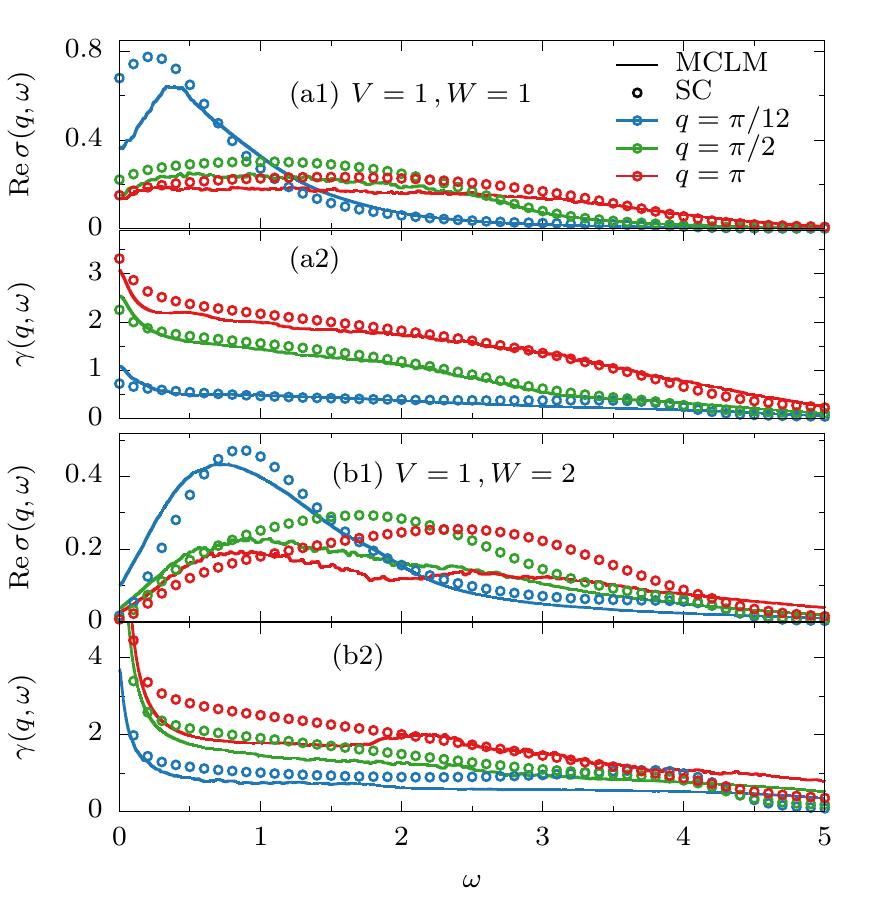}
\caption{(Color online) (a1,b1,c1) Dynamical conductivity $\mathrm{Re}\,\sigma(q,\omega)$ 
and (a2,b2,c2) current relaxation-rate function $\gamma(q,\omega)$, obtained via MCLM on 
$L=24$ sites for fixed parameters $V=1$, as compared to the solution of SC equations
(with effective length $L^*=24$), for different $q$ and two disorders: (a) $W=1$, and (b) $W=2$.}
\label{fig1}
\end{figure}

Characteristic results obtained for $L =24$ and averaged over $N_s \sim 100$ random 
configurations are presented in Fig.~\ref{fig1}. 
Some generic features be inferred: (a) Consistent 
with previous calculations of $\sigma(\omega)$ 
\cite{barisic10,gopal15,steinigeweg15,barisic16} our results indicate (for all disorders 
$W$) the maximum at $\omega = \omega^* >0$, and more important a nonanalytical 
low-$\omega$ behavior, i.e., 
$\sigma(\omega) \sim \sigma_0 + \zeta |\omega|^\alpha$. Here our
numerical results in the ergodic regime, $W<W_c $ imply an 
interpretation with $\sigma_0 >0$ and $\alpha \sim 1$ 
\cite{barisic10,steinigeweg15,barisic16}, while we comment later-on the possibility of 
the subdiffusion with $\sigma_0=0$ and $\alpha <1$ 
\cite{agarwal15,gopal15,znidaric16,luschen16}. (b) Within our resolution $\sigma_0$, but 
also general $\sigma(q,\omega \to 0)$, vanishes for $W\geq W_c$ consistent with the 
onset of the MBL phase and nonergodicity at all $q$. This implies necessarily via 
Eq.~\eqref{sigqw} a divergent $\gamma(q,\omega \to 0) \to \infty$, as also evident on 
approaching the MBL transition.

\begin{figure}[!htb]
\includegraphics[width=1.0\columnwidth]{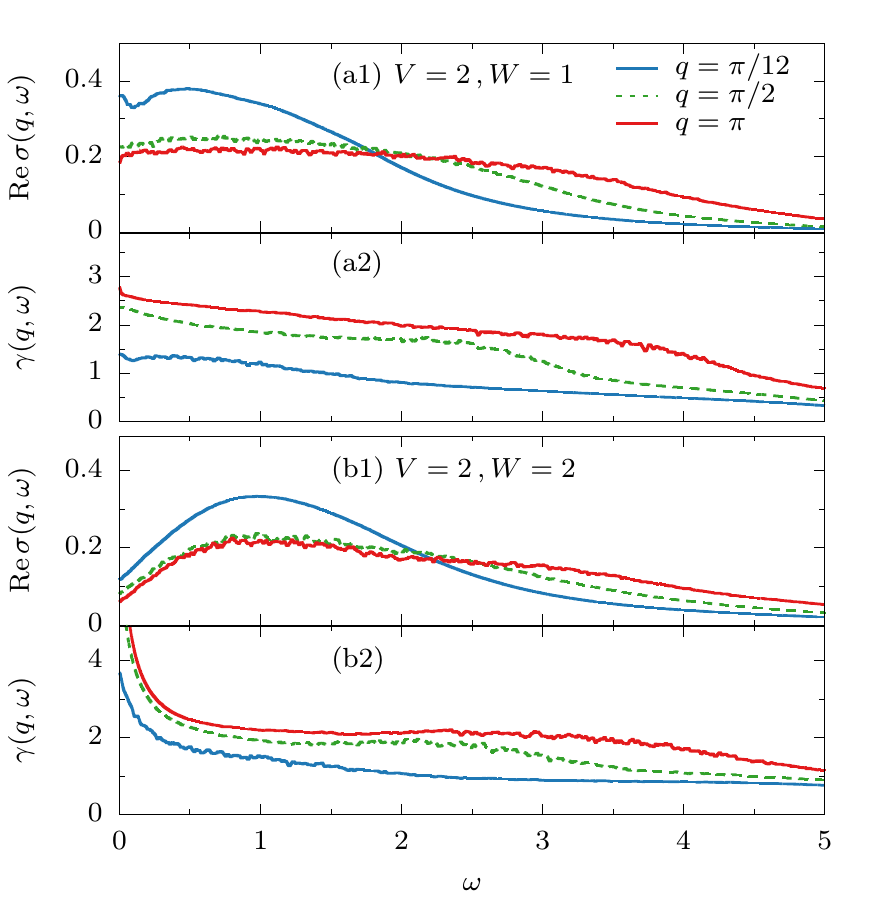}
\caption{(Color online) (a1,b1) $\mathrm{Re}\,\sigma(q,\omega)$ and (a2,b2) 
$\gamma(q,\omega)$ (MCLM, $L=24$) for fixed $V=2$, few different $q$, and 
two disorders: (a) $W=1$ and (b) $W=2$.}
\label{figS2}
\end{figure}

For comparison we present in Fig.~\ref{figS2} also corresponding numerical results for 
$V=2$, which corresponds to the isotropic Heisenberg model with random 
magnetic fields and has been in this connection studied more frequently. One can notice 
that larger $V$ does not change qualitatively results for both $\sigma(q,\omega)$ as well as 
$\gamma(q,\omega)$, but rather additionally broadens spectra, except the MBL 
singularity at $\omega \sim 0$ for $W=2$.

\section{Analytical approach to current decay-rate function}

\subsection{Effective force}

The motivation for following analytical approach and approximations comes form the 
perturbation theory, which can be performed for weak disorder $W \to 0$ (and somewhat more 
delicate for $V \to 0$ ) on the level of the current decay-rate function 
$\Gamma(q,\omega)$, in analogy to the theory of current scattering mechanisms in 
simple metals \cite{gotze72}. Such a theory has been extended to the nontrivial problem of
Anderson localization by taking it beyond the perturbative approximation \cite{gotze79,
vollhardt80,vollhardt801} and we will partly follow an analogous treatment for the
MBL problem. 
 
The expression for $\Gamma(q,\omega)$ (see the formal derivation and Eq.~\eqref{gamqw} in 
the Appendix B) is the starting point for the analytical approximations. 
The current scattering mechanism is determined by the operator for the effective force 
$F_q=Q {\cal L} j_q$, with the Liouville operator ${\cal L} j_q=[H,j_q]$ and $Q$ representing 
the operator \cite{mori65,forster95} which projects into space 
perpendicular to $n_q$ (see Appendix~B for details). ${\cal L} j_q$ can be evaluated 
explicitly from the model \eqref{tvw},
\begin{eqnarray}
{\cal L} j_q &=& t g_q h^d_q 
- \frac{1}{\sqrt{L}} \sum_k g_k \epsilon_k h^k_{q-k} \nonumber \\
&-& \frac{V}{\sqrt{L}} \sum_k w_k n_k h^k_{q-k} + 2 t^2 g_q n_q \,,
\label{ljq}
\end{eqnarray} 
where $w_k = 2 \sin(3k/2)$ and we define also (Fourier transforms of) kinetic-energy, 
potential and next--nearest hopping terms, respectively, 
\begin{eqnarray}
h^k_q &=& - \frac{t}{\sqrt{L}} \sum_i \mathrm{e}^{\imath q(i+1/2)} 
[ c^\dagger_{i+1} c_{i} + \mathrm{H.c.} ] \,,\nonumber \\
h^d_q &=& - \frac{t}{\sqrt{L}} \sum_i \mathrm{e}^{\imath qi}
[c^\dagger_{i+1} c_{i-1} + \mathrm{H.c.}]\,,\nonumber\\
\epsilon_k &=& \frac{1}{\sqrt{L}} \sum_i \mathrm{e}^{\imath qi} \epsilon_i\,.
\label{hkdq}
\end{eqnarray}

In the evaluation of $F_q $ the last term in Eq.~\eqref{ljq} vanishes due to $Q n_q =0$. 
Other three terms remain unaffected by the action of $Q$ within $\beta \to 0$ 
limit. With such a force operator $F_q$ we can write 
\begin{eqnarray}
\Gamma(q,\omega) &=& \frac{1}{\chi^0_j} \Lambda(q,\omega)\,, \nonumber \\
\Lambda(q,\omega) &\sim& \frac{\chi_F(q,\omega)-\chi^0_F(q)}{\omega}\,,
\label{lamqw}
\end{eqnarray}
where $\chi_F(q,\omega)$ are the generalized (force) susceptibilities, defined for the 
operator $F_q$ [compare with Eq.~\eqref{chiqw}]. In the above expression, in analogy 
to weak scattering theory \cite{gotze72,forster95}, we have introduced the 
(straightforward with an perturbative approach) approximation
neglecting the projections onto $n_q$ and $j_q$ space, 
${\cal L}_{ QQ^\prime} \to {\cal L}$, in the 
resolvent of Eq.~\eqref{lamqw} [compare Eq.~\eqref{gamqw}].

\subsection{Perturbative approximation}

Following Eq.~\eqref{fq} in the Appendix~B, we are dealing with force $F_q=F_{q1}+F_{q2}+F_{q3}$, 
representing different current scattering mechanisms. Similarly as in the derivation of 
the dynamical conductivity in metals \cite{gotze72}, we represent 
$\Gamma(q,\omega)$ as the sum of three contributions, neglecting possible mixed 
correlations. 

$\Lambda_1(q,\omega) \propto g_q^2$ vanishes in the hydrodynamic regime $q \to 0$ and 
can be approximated by the NI limit, i.e.,
\begin{equation}
\Gamma_1(q,\omega)= \frac{g_q^2}{2} \frac{1}{L} \sum_k \frac{- f_{k}^2} {\omega^+ + 
e_{k+q/2} -e_{k-q/2}}\,,
\label{gam1}
\end{equation}
where $f_k=e_{2k}$ and $e_k =-2t \cos k$ is the NI fermion dispersion. It should be, however,
 recognized that even at $W =V=0$ the $\Gamma_1(q,\omega)$ does not fully reproduce the NI result 
for $\phi(q,\omega)$ [due to simplification in Liouville operator in 
Eqs.~\eqref{gamqw},\eqref{lamqw}]. This deficiency is easily remedied by noticing that the correct NI 
result is obtained by replacing $f_k = e_{k}$ in Eq.~\eqref{gam1}. Still, for $W>0$ 
single-particle eigenstates do not have a well defined wavevector $k$, so more 
reasonable approximation in this case is to assume an additional broadening, 
i.e., $\delta = W/\sqrt{3}$, corresponding to the width of the random-potential
distribution. This details hardly influence any qualitative results further-on, 
since $\Gamma_1(q,\omega)$ does not contribute in the hydrodynamic regime $q \to 0$.

By decoupling the static disorder and dynamical density fluctuations in 
$\Gamma_2(q,\omega)$, we get
\begin{equation}
\Lambda_2(q,\omega)= \frac{1}{L} \sum_k g_{q-k}^2
\langle \epsilon_{q-k}^2 \rangle \phi_{k}(k,\omega)\,,
\label{lam2}
\end{equation}
where $\phi_{k}(q)$ is the relaxation function of the kinetic-energy $h^k_q$, Eq.~\eqref{hkdq}, 
defined in analogy to $\phi(q,\omega)$, Eq.~\eqref{chiqw}. 
In the NI limit (but with disorder $W>0$) Eq.(\ref{lam2}) reduces to 
\begin{equation}
\Gamma_2(q,\omega)=\frac{W^2}{6 t^2 L^2} \sum_{k,k'} \frac{- g_{q-k}^2 e^2_{k'} }
{\omega^+ + e_{k'+k/2} -e_{k'-k/2}}\,,
\label{gam2}
\end{equation}
which is the lowest-order scattering (Boltzmann-type) result \cite{gotze72}, in 
particular it gives a finite relaxation rate 
$\gamma(q,\omega) = \mathrm{Im}\,\Gamma(q,\omega)$, also in the 
hydrodynamic $(q, \omega)\to 0$ limit. 
 
The perturbative treatment of the interaction term is more problematic. One can assume that 
the dynamical fluctuations of density $n_k$ and kinetic energy $h^k_{q-k}$
are independent, which leads to 
\begin{eqnarray}
\mathrm{Im}\,\Lambda_3(q,\omega) &=& \frac{V^2}{L} \sum_k \frac{w_k^2}{\pi \beta} 
\int_{-\infty}^{\infty} \mathrm{d}\omega^\prime \times \nonumber \\
&&\mathrm{Im}\,\phi(k,\omega^\prime) \mathrm{Im}\,\phi_{k}(q-k,\omega-\omega^\prime)\, .
\label{lam3}
\end{eqnarray}
When we insert the NI input for $\phi(q,\omega)$
and $\phi_{k}(q,\omega)$, the interaction $V>0$ leads to additional current decay channel, even at
$(q , \omega) \to 0$. While this is an effect generally expected from the inter-particle 
interaction, in our particular case it is not fully justified since the pure ($W=0$) model is 
integrable and exhibits a dissipationless current and singular 
$\sigma(\omega \sim 0) = \beta D \delta(\omega)$ 
with $D>0$ even at $\beta \to 0$. Since we are interested more in the role
of disorder and of a generic interaction term, where current dissipation should emerges from a 
term like Eq.~\eqref{lam3}, we would here stay at this level of approximation. 

\subsection{Self-consistent closure}

At this stage we are not aiming to develop more detailed theory for 
kinetic-energy fluctuations $\phi_{k}(q,\omega)$ entering Eqs.~\eqref{gam2},\eqref{lam3}. It is, however, crucial 
to take into account the fact that the kinetic-energy function has an overlap 
with the energy-density relaxation function. In a disordered system, the energy is, 
besides the number of particles, the only conserved quantity. It is therefore 
essential to take properly the $q \to 0$ energy fluctuations, and we treat these 
correlations in analogy to Eqs.~\eqref{mem},\eqref{sigqw} with the role of 
$\sigma(q,\omega)$ replaced by the thermal conductivity $\kappa(q,\omega)$. The latter 
has been found \cite{karahalios09,prelovsek17} to have similar behavior close to the MBL 
transition, in particular the vanishing of d.c. value $\kappa_0$ and anomalous 
low-$\omega$ behavior. Taking into account sum rules 
$\eta= \chi^0_{k}/\chi^0 =2 t^2$ we further-on work with a simplification 
$\phi_{k}(q,\omega)= \eta \phi(q,\omega)$ representing an effective Wiedemann-Franz 
relation, i.e., assuming the same relaxation rates for density and energy currents.

Since $\Lambda \propto \phi \propto \beta$, we further-on work with renormalized 
relaxation functions, i.e., $\widetilde \phi=\phi/\chi^0, \widetilde \phi_{k}=\phi_{k}/\chi^0$. 
So the final SC equations, 
besides $\Gamma_1(q,\omega)$, where we do not correct Eq.~\eqref{gam1},
are 
\begin{eqnarray}
\Gamma_2(q,\omega) &=& \frac{\eta W^2}{ 6 t^2 L} \sum_k
g_{q-k}^2 \widetilde{\phi}(k,\omega)\,, \label{sgam2} \\
\mathrm{Im}\,\Gamma_3(q,\omega) &=& \frac{\eta \widetilde{n}V^2 }{2 \pi L} 
\sum_k w_k^2 \int \mathrm{d}\omega^{\prime}\,
\mathrm{Im}\,\widetilde{\phi}(k,\omega^{\prime}) \times \nonumber \\
&&~~~~~~~~~~ \mathrm{Im}\,\widetilde{\phi}(q-k,\omega-\omega^{\prime})\,,
\label{gam3}
\end{eqnarray}
where $\widetilde{n}=\bar n(1-\bar n)$. The MBL 
physics, in particular the transition, is predominantly governed by 
$\Gamma_2(q,\omega)$, while for $W>0$ the interaction-driven term 
$\Gamma_3(q,\omega)$, due to convolutions in $(q,\omega)$, yields rather a featureless 
function, leading to the current decay at all $q$.
 
Due to the coupling to the $q \to 0$ diffusion mode in 
Eq.~\eqref{gam1}, it is evident that $\Gamma(q,\omega \to 0)$ as well as the whole SC 
set might be singular in 1D. In order to simulate finite-size systems (as studied 
numerically) and explore the finite-size scaling we introduce a finite cutoff 
$k_m= \pi /L^*$, in particular in Eq.~\eqref{sgam2}. It should be pointed out that after 
taking mentioned simplifications there are (at given model constants $V,W$) no free 
parameters in the SC theory apart from the cutoff $k_m$ (effective length $L^*$).

We note that presented SC equations have an analogy to simplified theories of Anderson 
localization \cite{gotze79}. It has been, however, established that proper SC 
localization theory for NI fermions \cite{vollhardt80,vollhardt801} should 
take into account the time-reversal symmetry of correlation functions on a 
single-particle level. The latter is, however, lost by including finite interaction 
$V>0$. As a consequence, Eq.~\eqref{sgam2} emerges as a nontrivial coupling of 
only remaining low-$\omega$ collective modes in the system, i.e., the density and the energy diffusion 
mode. 

\section{Numerical solutions of SC equations}

\subsection{General features}

Having SC set of equations, Eqs.~\eqref{mem},\eqref{sigqw},\eqref{gam1},\eqref{sgam2},\eqref{gam3}, 
it is straightforward to find solutions by numerical 
iteration of coupled equations until convergence, whereby we use at the initial step the 
NI input for $\phi(q,\omega)$. In Fig.~\ref{fig1} we present typical 
SC-theory results for $\mathrm{Re}\,\sigma(q,\omega)$ and $\gamma(q,\omega)$ along with 
the MCLM numerical ones, whereby we use $L^*=24$ corresponding to the size used in MCLM
calculation. Qualitative agreement is quite satisfactory for modest value of disorder strength $W$, in 
particular the analytical theory reproduces some essential features: (a) Maxima of 
$\sigma(q,\omega)$ with $\omega^* >0$ emerge also in SC solutions due to a nontrivial 
maxima in $\gamma(q,\omega \to 0)$. The maximum moves towards $\omega^* \sim 0$ for weaker disorder 
$W < 0.8$, which would be the signature of a normal diffusion. (b) In the ergodic regime at $W < W_c^* \sim 1.6$ 
low-$\omega$ SC results for small $L^* < 100$ can be roughly fitted to
$\sigma(\omega)=\sigma_0 + b |\omega|^\alpha$ with $\sigma_0 > 0$ and 
$\alpha \sim 1$ close to the MBL transition. (c) Due to a large increase of $\gamma(q,\omega \to 0)$ the 
conductivity $\sigma(q,\omega \to 0)$ is strongly reduced for larger $W >1$. (d) 
Eventually, for $W > W_c^*$ the SC equations yield a singular solution 
$\gamma(q,\omega \sim 0)= \gamma_s \delta(\omega)$ which is the hallmark of the 
nonergodicity and leads also to vanishing d.c. transport $\sigma(q,\omega \to 0) = 0$.

\begin{figure}[!htb]
\includegraphics[width=1.0\columnwidth]{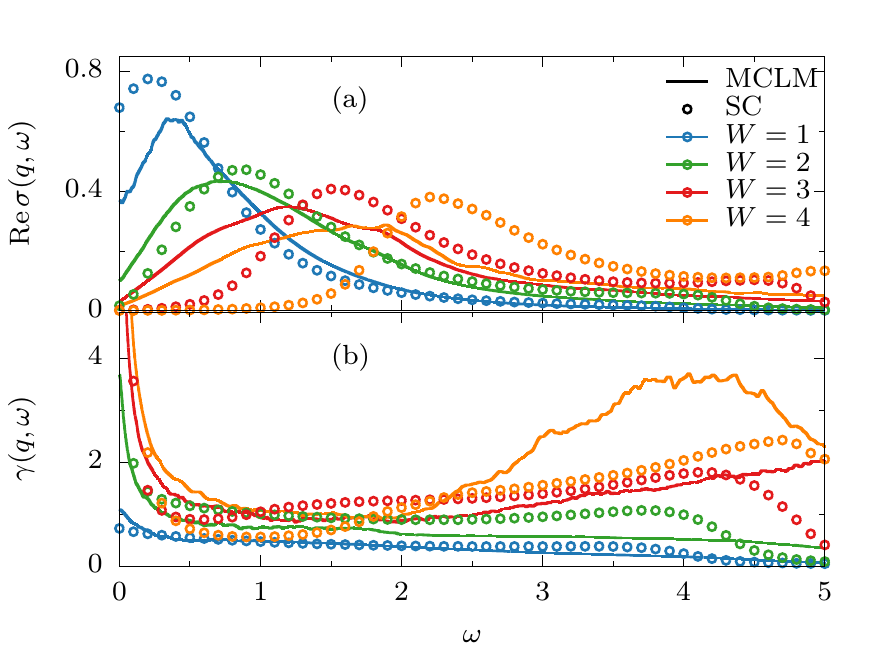}
\caption{(Color online) Comparison between SC and numerical (MCLM) results of (a) 
$\mathrm{Re}~\sigma(q,\omega)$ and (b) $\gamma(q,\omega)$ for $q=\pi/12$, 
$L=L^*=24$, and various disorder strengths $W=1,2,3,4$.}
\label{figS22}
\end{figure}

The behavior with $W$ varying across the MBL transition is presented in Fig.~\ref{figS22}, where we
compare results for disorder strength up to $W=4 \gg W_c^*$. We observe that the quantitative agreement between 
SC and numerical result is steadily  decreasing with increasing $W > W_c^*$. 
This coincides with the fact that SC threshold $W_c^* \sim 1.6$ is significantly below 
numerical (at $V=1$) estimate $W_c \sim 3$ ~\cite{lev15}. The origin of this discrepancy
in critical $W_{c}^*$ can traced back to overestimated coupling between density and energy 
diffusion mode enhancing the feedback (localization) mechanism in SC equations via the  
$\gamma(q,\omega \to 0)$ behaviour. Still, the overall qualitative change across the MBL
transition follows the same pattern as the numerical one.

\begin{figure}[!htb]
\includegraphics[width=1.0\columnwidth]{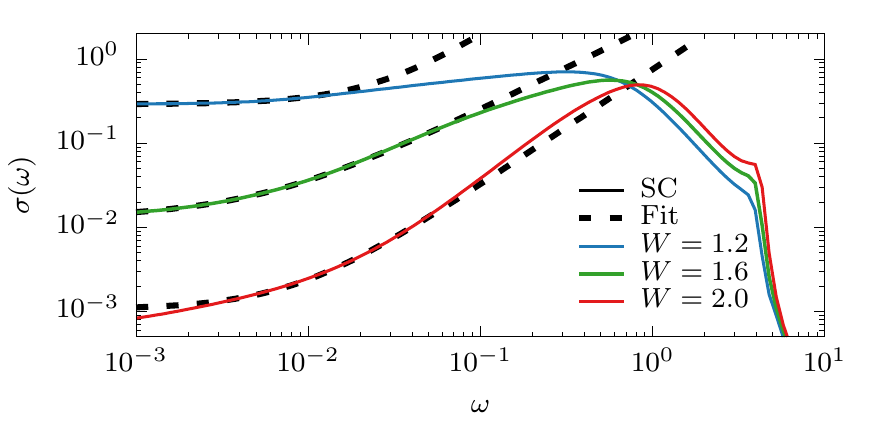}
\caption{(Color online) Comparison of SC solution (solid line, $L^*=40$, 
$\delta\omega=10^{-3}$) with the fit $\tilde \sigma(\omega)=a+b|\omega|^c$ (dashed line). 
with $c= 0.8,1.0,1.5$ for $W=1.2,1.6,2.0$, respectively.}
\label{figS33}
\end{figure}

When we are comparing SC results for optical conductivity $\sigma(\omega) = \sigma(q \to 0, \omega)$
with previous numerical studies (as well as this study for $q>0$) on finite systems \cite{karahalios09,steinigeweg15,barisic16}, 
we should use appropriate  $L^*$ as well as corresponding $\delta \omega$.  
In Fig.~\ref{figS33} we present a characteristic result for modest $L^*=40$ (and $\delta \omega \sim 10^{-3}$)
for $\sigma(\omega)$ across the transition to the MBL, i.e. $1.2 \leq W \leq 2.0$, 
together with the low-$\omega$ fit to $\tilde \sigma(\omega)=a+b|\omega|^c$. 
We note that such a fit should be evidently restricted to the range well below the maximum 
$\omega \ll \omega^*$ which is for $W > W_c^*$ at $\omega^* \sim 1$, but for lowest $W=1.2$ moves
down to $\omega^* \sim 0.2$\cite{steinigeweg15}. Nevertheless, the overall behavior around the transition $W \sim W_c^*$ is
is characterized by $\alpha \sim 1$ and a clear drop of $\sigma_0$.

\subsection{Subdiffusion and transition to MBL}

\begin{figure}[!htb]
\includegraphics[width=1.0\columnwidth]{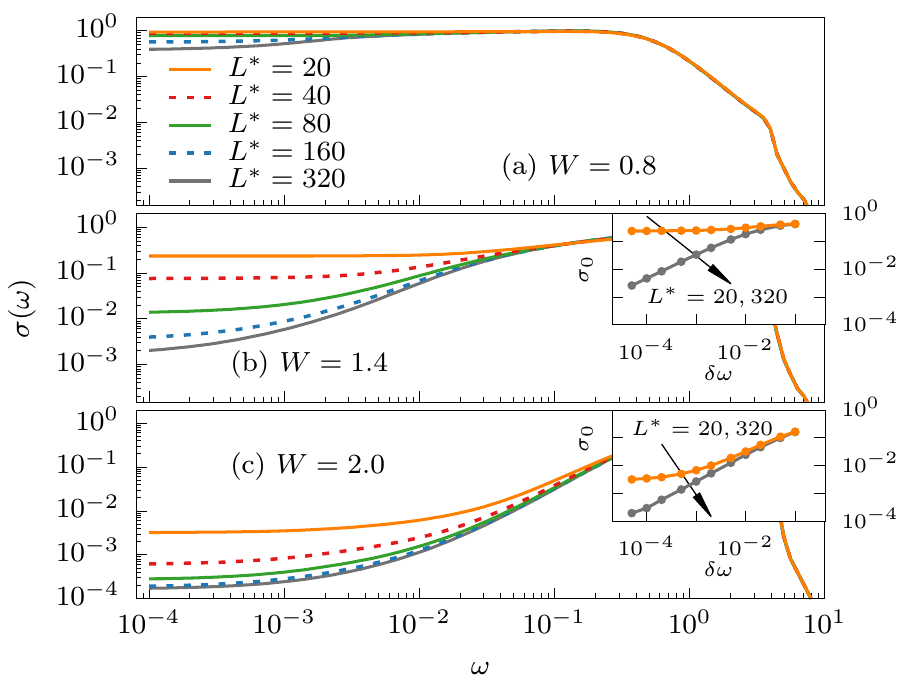}
\caption{(Color online) Optical conductivity $\sigma(\omega)$ (in a log-log 
scale) as evaluated from the SC theory at $V=1$ for different $W=0.8, 1.4, 2.0$, and
various effective lengths $L^* = 20 - 320$ with frequency resolution $\delta \omega=10^{-4}$.
Insets of (b,c): scaling of $L^*=20$ and $L^*=320$ with $\delta \omega$ used in SC equations.}
\label{fig2}
\end{figure}

\begin{figure}[!htb]
\includegraphics[width=1.0\columnwidth]{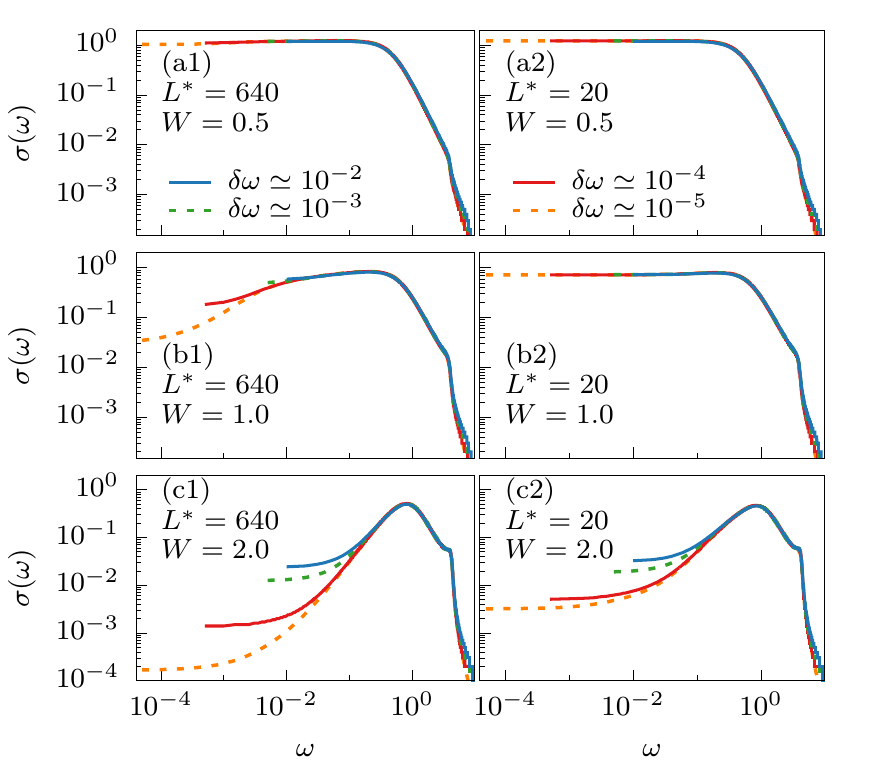}
\caption{(Color online) (a-c) Frequency resolution $\delta\omega$ dependence of 
$\sigma(\omega)$ for fixed effective length $L^*=640$ (left column) and $L^*=20$ (right column), 
and various disorder strength $W=0.5,1.0,2.0$.}
\label{figS3}
\end{figure}

While SC results in Fig.~\ref{fig1} (as well as Fig.~\ref{figS33}) show an overall behavior 
for $W<W_c^*$ and $W \sim W_c^*$, 
consistent with numerical results at finite $L^*$, we further investigate in more detail the 
consequences of the singular aspects due to 1D. In order to explore the low-$\omega$ 
behavior, we concentrate on most interesting $q \to 0$ results and present in 
Fig.~\ref{fig2} $\sigma(\omega)$ as obtained with large frequency $\omega$ resolution 
($\delta \omega \sim 10^{-4}$) at several characteristic $W$ and varying effective 
length $L^* = 20 - 320$. 
It should be realized that the choice of $\delta \omega$ in the numerical
SC procedure is intimately related to $L^*$ and we cannot get strictly 
$\sigma_0 =0$ at $\delta \omega>0$. Nevertheless the scaling $\delta \omega \to 0$,
as shown in the inset of Fig.~\ref{fig2}, is consistent with vanishing 
$\sigma_0=0$, at least for $W > 1.2$. This is also presented in Fig.~\ref{figS3}, which
depicts dependence of dynamical conductivity $\mathrm{Re}\,\sigma(q,\omega)$ on 
frequency resolution, $\delta\omega$, for fixed cutoff $L^*=20$ and $L^*=640$ and 
various disorder strength.

Taking this into account, we can distinguish three regimes as already 
noted in numerical studies \cite{agarwal15,gopal15,znidaric16}: (a) At small disorder 
$W <1 $ $\sigma(\omega \to 0)$ is only weakly dependent on $L^*$ and it is hard to 
detect signatures of a subdiffusion even at extreme $L^* \gg 100$. (b) At the intermediate 
$1 < W < W^*_c$ we confirm the steady decrease of $\sigma_0$ with increasing $L^*$ and the 
behavior can be well captured with subdiffusion form
$\sigma(\omega) \propto |\omega| ^\alpha$ with $\alpha<1$. (c) For $W>W^*_c$ results 
become again only weakly $L^*$ dependent, while the d.c. value $\sigma_0$ is vanishing.

\begin{figure}[!htb]
\includegraphics[width=1.0\columnwidth]{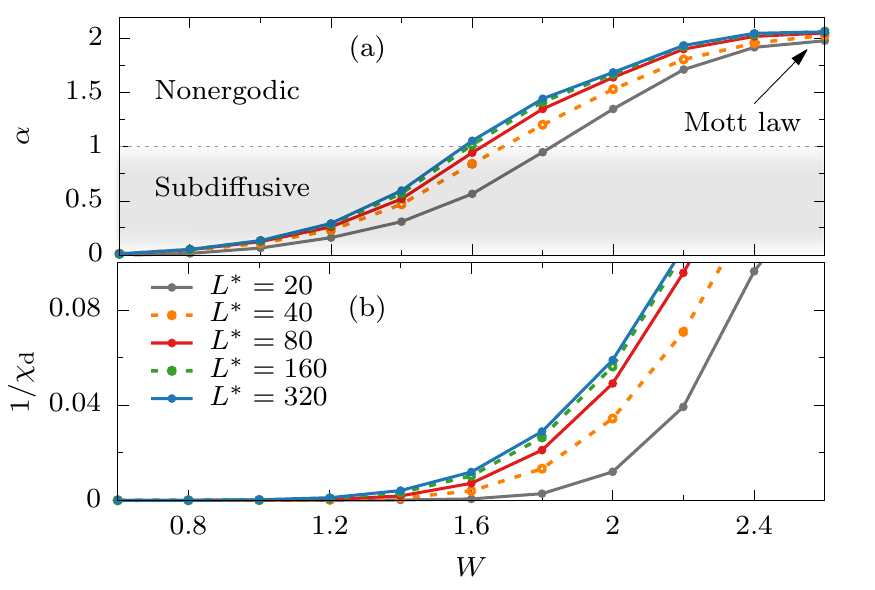}
\caption{(Color online) (a) Dynamical exponent $\alpha$ and (b) the inverse dielectric 
polarizability $1/\chi_{\mathrm{d}}$ vs. $W$ as evaluated at $V=1$ and different 
$L^*=20 - 320$ and $\delta \omega=10^{-4}$. Note that the MBL transition is determined by $\alpha=1$.}
\label{fig3}
\end{figure}

To make the analysis of subdiffusion more objective, we define the exponent via the 
maximum slope 
\begin{equation}
\alpha = \frac{ \mathrm{d}\,\log\sigma(\omega)}{ \mathrm{d}\,\log \omega}\,,
\label{alpha}
\end{equation}
 in the 
range $\omega<0.1$. Results are shown in Fig.~\ref{fig3}(a). It is indicative that the 
subdiffusion with $\alpha \ll 1$ can be hardly established for $W <1$ since it requires 
$L^* \gg 100$~\cite{znidaric16}. On the other hand, results with $\alpha > 0.3$ are 
better resolved. The crossing $\alpha =1$ marks the MBL transition to the nonergodic 
phase, where for large $W \gg W^*_c$ we get $\alpha \sim 2$, as expected deep inside 
the localized regime \cite{mott68}.

As an uniform ($q \to 0$) order parameter 
within the MBL (nonergodic) phase one can consider the current-relaxation stiffness 
$\gamma_s(q) >0$. More physical is the dielectric polarizability 
\begin{equation}
\chi_{\mathrm{d}} = \frac{2}{\pi }\int_{0}^{\infty} \frac{\sigma(\omega)}{\omega^2}
\mathrm{d}\omega\,,
\end{equation} 
whereby $\chi_{\mathrm{d}} < \infty $ implies that the system is dielectric, i.e., an 
external field along the chain induces only a finite polarization. It is evident, that 
$\alpha> 1$ is required for $\chi_{\mathrm{d}} <\infty$. In Fig.~\ref{fig3}(b) we 
present results for the inverse $1/\chi_{\mathrm{d}}$ vs. $W$ as evaluated for different 
$L^*$, revealing indeed its vanishing below the MBL transition.

\section{Conclusions}

Presented analytical theory is a SC extension of the perturbative evaluation of the 
current decay-rate function $\Gamma(q,\omega)$. The disorder effect is reproduced within 
the lowest order (Boltzmann-type) scattering, while the interaction is treated only 
within a decoupling approximation. In analogy to the SC theory of the single-particle 
Anderson localization \cite{gotze79,vollhardt80,vollhardt801}, the theory is closed 
beyond the weak scattering approximation, where the crucial assumption (at the present 
level of the theory) is that density and energy dynamical correlations are related, in 
particular at small $(q,\omega)$, and both simultaneously undergo a MBL transition. 

Although the theory starts from the lowest-order calculation of the current-relaxation
function $\Gamma(q,\omega)$ its extention into a SC scheme goes well beyond 
the perturbative approach. A SC determination of $\Gamma(q,\omega)$ leads, on approaching 
the MBL transition, to enhanced low-$\omega$ density and energy-density fluctuations,
which finally leads to the freezing of the low-$\omega$ dynamics at the $W=W^*_c$. 
Beyond this disorder correlation functions are nonergodic and characterized by singular
contribution in $S(q,\omega) \sim S_q \delta(\omega)$ with $S_q > 0$. 
One might question the particular validity and the form of the SC loop, however the freezing of 
low-$\omega$ dynamics is well visible in numerical results and consistent with analogous 
phenomenon  in the theory of Anderson localization \cite{gotze79,vollhardt80,vollhardt801}.

In the presented theory there are no free parameters except the cutoff $k_m=\pi/ L^*$  which 
simulates the finite-size system and allows for the finite-size scaling. The importance 
of cutoff and corresponding sensitivity of SC solutions on the frequency resolution 
$\delta \omega$ appears to be a singular property of 1D and makes the proper convergence
of solutions of coupled analytical equations nontrivial.
In spite of simplifications the presented SC theory yields several nontrivial conclusions, 
consistent with numerical results obtained in this paper via the MCLM method, 
but also with previous numerical investigations on finite systems:

\noindent (a) When simulating numerically reachable finite-size systems by taking 
cutoff $L^* \sim 20-40$  (as well corresponding finite-frequency
resolution $\delta \omega \sim 10^{-3}$) our SC results appear to be consistent
with the dynamical conductivity $\sigma(\omega) \sim
\sigma_0 + b |\omega|^\alpha$ with $\alpha \sim 1$ and vanishing $\sigma_0$ near the 
MBL transition \cite{karahalios09,steinigeweg15,barisic16}.

\noindent (b) However, careful scaling beyond $L^* > 100$ and $\delta\omega\ll 10^{3}$
of SC solution indicates  on vanishing $\sigma_0=0$ in the ergodic regime $W<W_c^*$ 
(at least for $W>1.2$ at $V=1$). Within the present SC theory this emerges
due to the disorder-induced coupling between the density and the energy 
diffusion mode. As a consequence of 1D, in the ergodic regime the transport is 
subdiffusive \cite{agarwal15,gopal15,luitz16,luitz116,luschen16}, i.e., for large enough systems 
d.c. transport coefficients are expected to vanish, e.g., $\sigma_0 \to 0$. Still, for modest 
disorder $W$ effective sizes to detect such anomalies could be huge, e.g. $L^* \gg 100$ 
\cite{znidaric16,luitz116}, and therefore hard to detect in numerical and even 
experimental studies.

\noindent (c) The transition to the nonergodic MBL regime $W > W_c$ appears in the theory 
via the onset of the current-decay stiffness $\gamma_s >0$, which coincides with the 
condition for the dynamical exponent $\alpha >1$ and the dielectric 
polarizability $\chi_{\mathrm{d}} <\infty$. 

\noindent (e) Theoretical 
results for dynamical correlations show an overall qualitative agreement with numerical 
ones (at corresponding effective length $L^*$) in the whole $(q,\omega)$ range.

When we discuss the validity and restrictions within the presented theory, there are several 
aspects in which should be considered:
 
\noindent (a) Since the theory is an extension
of the perturbative treatment of disorder starting at modest $W$, it is plausible that 
we cannot claim a quantitative agreement for larger disorder with $W \sim W^*_c$ or even 
more within the MBL regime $W> W^*_c$. The reason is mainly twofold: At $W > 3$ 
single-particle states are already well localized. Still, more problematic seems to be the 
overestimated coupling between  density and energy  diffusion modes, which leads to 
overestimated feedback in SC equation and consequently to the transition at
critical $W_c^*$, substantially smaller than emerging from numerical studies 
(e.g. at $V=1$ $W_c^* \sim 1.6$ instead of numerical estimate $W_c=3$). 
 This can be improved by taking both relevant hydrodynamic modes,
i.e., density and energy diffusion, on equal footing into the analysis. In this work we skip 
this aspect in order to make our SC theory as transparent as possible. 

\noindent (b) The current decay-rate due to interaction $V>0$ is taken very crudely,
in particular since the actual model without disorder (at $W=0$) is integrable and 
$V >0$ itself does lead to d.c. conductivity $\sigma_0 < \infty$. Nevertheless, generic
interaction term is expected to lead to the scattering of d.c. current (at $T \gg 0$). Moreover 
$\Gamma_3(q,\omega)$ seems to be less critical dependent on dimensionality of the system, as appears the
case for $\Gamma_2(q,\omega)$ emerging from disorder. 

\noindent (c) The assumption that
the dynamical quantities are self-averaging is inherent in the SC approach,
although this aspect should be further critically examined due to possible
role of rare large disorder fluctuations \cite{agarwal15,gopal15}. 

The presented SC scheme is more generic and can be generalized into different 
directions. Analogous treatment of higher dimension in rather straightforward, 
especially since some anomalies as, e.g., the subdiffusion are not expected there, at least
not to such extent. One could treat also separately the density and energy dynamical 
correlations, whereby the latter one are much less investigated so far. On the other hand, 
for experiments on MBL in 
cold-atom systems \cite{schreiber15,bordia16,luschen16} the relevant model is the 
disordered Hubbard model which does reveal a disorder-induced spin-charge separation 
\cite{prelovsek16}, which might also be approached in a similar way.

\begin{acknowledgments}
P.P. acknowledges the support by the program P1-0044 of the Slovenian Research Agency. 
J.H. acknowledges the support by the U.S. Department of Energy, Office of Basic Energy 
Sciences, Materials Science and Engineering Division.
\end{acknowledgments}

\section*{Appendix A: Correlation functions} 

Since the system under consideration can be nonergodic, one should be careful with 
the definitions of correlation and response functions. In our analysis we define the 
dynamical susceptibility (response functions) $\chi_A(\omega)$ and corresponding static 
(thermodynamic) response $\chi_A^0$, for arbitrary operators $A$ in the standard way,
\begin{eqnarray}
\chi_{A}(\omega) &=& -\imath\int_0^\infty\mathrm{d}t\, \mathrm{e}^{i \omega^+ t}
\langle[A^\dagger(t),A]\rangle\,, \\
\chi^0_{A} &=& \int_0^{\beta}\mathrm{d} \tau\, \langle A^\dagger A(i \tau) \rangle =
(A|A)\,,
\label{chia}
\end{eqnarray}
where $\omega^+ = \omega + i \delta$ with $\delta \to 0$ and $\beta =1/T$. 
Eq.~\eqref{chia} introduces the scalar product \cite{mori65,forster95},
convenient for formal representation and derivation of memory functions, 
even for nonergodic systems. 
Above $\langle ..\rangle$ denotes the canonical thermodynamical 
average and in a disordered system additional averaging over all random 
configurations of $\epsilon_i$ (see the comment in the main text after 
Eq.~(\ref{chiqw})).

In the analysis, instead of susceptibilities $\chi_A(\omega)$, we mostly use related 
relaxation functions,
\begin{equation}
\phi_A(\omega)= \frac{\chi_{A}(\omega)-\chi^0_{A}}{\omega}
=(A|\frac{1}{{\cal L}-\omega}|A)\,,
\label{phia}
\end{equation}
where the second representation in 
Eq.~\eqref{phia} in terms of the resolvent with Liouville operator $
{\cal L} A = [H,A]$ is a standard one allowing formal steps further-on. 
The nonergodic behavior is in this framework 
characterized by the behavior $\chi^0_A > \chi_A(\omega \to 0)$ leading to a singular 
low-$\omega$ contribution \cite{pirc74,gotze79,vollhardt80,vollhardt801},
\begin{equation}
\mathrm{Im}\,\phi_A(\omega \sim 0) = \pi D_A \delta(\omega)\,,
\qquad D_A = \chi^0_{A} -\chi_{A}(\omega \to 0)\,,
\end{equation}
where $D_A$ is the corresponding stiffness. 

Finally, since we are dealing only with the case of high-$T$, i.e. $\beta \to 0$, there are 
convenient simplification following from Eq.~\eqref{chia} and Eq.~\eqref{phia}
\begin{equation}
\chi^0_A= \beta \langle A^\dagger A\rangle, \quad \phi_{A}(\omega) =
-\imath \beta \int_0^\infty \mathrm{d}t\,\mathrm{e}^{\imath \omega^+ t} 
\langle A^\dagger (t) A\rangle\,,
\label{beta0}
\end{equation}
and in particular simplified relation to the general dynamical structure factor 
$\mathrm{Im}\,\phi(\omega) = \pi \beta S_A(\omega)$.

\section*{Appendix B: Memory-function representation}

We use definitions above for several operators $A$ of interest. The starting point 
are density relaxation function with $A=n_q$,
The memory function (MF) representation of $\phi(q,\omega)$ follows from the 
continuity equation,
\begin{equation}
{\cal L} n_q= g_q j_q\,, \quad j_q = \frac{t}{\sqrt{L}} \sum_i
\mathrm{e}^{\imath q(i+1/2)}( \imath c_{i+1}^\dagger c_i +\mathrm{H.c.} )\,,
\end{equation}
where $ g_q = 2\sin(q/2)$. By defining the projection projector $P$
and its complement $Q$, 
\begin{equation}
P= |n_q) \frac{1}{\chi^0(q)} (n_q|\,, \qquad Q=1-P\,,
\end{equation}
where $\chi^0(q)=(n_q|n_q)$, we can express relaxation 
function, Eq.~\eqref{phia}, in the from of MF representation
\begin{equation}
\phi(q,\omega)= \frac{- \chi^0(q)}{\omega+ 
\imath g_q^2 \sigma(q,\omega)/\chi^0(q)}\,, 
\end{equation}
with 
\begin{equation}
\sigma(q,\omega)= (Q j_q| \frac{ - \imath}{{\cal L}_Q-\omega}| Q j_q) 
= (j_q| \frac{ -\imath}{{\cal L}_Q-\omega}| j_q)\,,
\label{ssigqw}
\end{equation}
where ${\cal L}_Q = Q {\cal L} Q$ is projected Liouville operator and $Q j_q=j_q$ by 
symmetry. It should be noted that $\sigma(q,\omega)$ is in general not equal to 
standard conductivity $\tilde \sigma(q,\omega)$, evaluated directly replacing in 
Eq.~\eqref{ssigqw} the reduced dynamics with the full one, ${\cal L}_Q \to {\cal L}$. 
Still, both quantities merge in the hydrodynamic limit $q\to 0$~\cite{forster95,gotze72}.

In the next step we express $\sigma(q,\omega)$ in terms of the current relaxation-rate 
function $\Gamma(q,\omega)$,
\begin{equation}
\sigma(q,\omega)= \imath \frac{\chi^0_{j}(q)}{\omega+\Gamma(q,\omega)}\,,
\end{equation}
where $\chi^0_{j}(q)=(j_q|j_q) $. While such a possibility follows directly from the 
analytical properties of $\phi(q,\omega)$ and $\sigma(q,\omega)$, the formal expression 
(used further-on as the starting point for analytical approximations in Sec.~IV ) can be given, 
introducing additional projector 
\begin{equation}
P^\prime =| j_q)\frac{1}{\chi^0(q)}(j_q|,\qquad Q^\prime=1-P^\prime\,,
\end{equation}
so that
\begin{equation}
\Gamma(q,\omega) = \frac{1}{ \chi^0_{j}(q) }
( F_q| \frac{1}{{\cal L}_{QQ^\prime}-\omega}| F_q) = 
\frac{1}{ \chi^0_{j}(q) } \Lambda(q,\omega)\,,
\label{gamqw}
\end{equation}
where (formally) ${\cal L}_{QQ^\prime}= Q^\prime {\cal L}_{Q} Q^\prime$ and 
\begin{equation}
F_q = Q Q^\prime {\cal L} j_q = Q {\cal L} j_q\,.
\label{fq}
\end{equation}

\bibliography{manuscmbl}

\begin{thebibliography}{45}%
\makeatletter
\providecommand \@ifxundefined [1]{%
 \@ifx{#1\undefined}
}%
\providecommand \@ifnum [1]{%
 \ifnum #1\expandafter \@firstoftwo
 \else \expandafter \@secondoftwo
 \fi
}%
\providecommand \@ifx [1]{%
 \ifx #1\expandafter \@firstoftwo
 \else \expandafter \@secondoftwo
 \fi
}%
\providecommand \natexlab [1]{#1}%
\providecommand \enquote  [1]{``#1''}%
\providecommand \bibnamefont  [1]{#1}%
\providecommand \bibfnamefont [1]{#1}%
\providecommand \citenamefont [1]{#1}%
\providecommand \href@noop [0]{\@secondoftwo}%
\providecommand \href [0]{\begingroup \@sanitize@url \@href}%
\providecommand \@href[1]{\@@startlink{#1}\@@href}%
\providecommand \@@href[1]{\endgroup#1\@@endlink}%
\providecommand \@sanitize@url [0]{\catcode `\\12\catcode `\$12\catcode
  `\&12\catcode `\#12\catcode `\^12\catcode `\_12\catcode `\%12\relax}%
\providecommand \@@startlink[1]{}%
\providecommand \@@endlink[0]{}%
\providecommand \url  [0]{\begingroup\@sanitize@url \@url }%
\providecommand \@url [1]{\endgroup\@href {#1}{\urlprefix }}%
\providecommand \urlprefix  [0]{URL }%
\providecommand \Eprint [0]{\href }%
\providecommand \doibase [0]{http://dx.doi.org/}%
\providecommand \selectlanguage [0]{\@gobble}%
\providecommand \bibinfo  [0]{\@secondoftwo}%
\providecommand \bibfield  [0]{\@secondoftwo}%
\providecommand \translation [1]{[#1]}%
\providecommand \BibitemOpen [0]{}%
\providecommand \bibitemStop [0]{}%
\providecommand \bibitemNoStop [0]{.\EOS\space}%
\providecommand \EOS [0]{\spacefactor3000\relax}%
\providecommand \BibitemShut  [1]{\csname bibitem#1\endcsname}%
\let\auto@bib@innerbib\@empty
\bibitem [{\citenamefont {Anderson}(1958)}]{anderson58}%
  \BibitemOpen
  \bibfield  {author} {\bibinfo {author} {\bibfnamefont {P.~W.}\ \bibnamefont
  {Anderson}},\ }\bibfield  {title} {\enquote {\bibinfo {title} {Absence of
  diffusion in certain random lattices},}\ }\href {\doibase
  10.1103/PhysRev.109.1492} {\bibfield  {journal} {\bibinfo  {journal} {Phys.
  Rev.}\ }\textbf {\bibinfo {volume} {109}},\ \bibinfo {pages} {1492} (\bibinfo
  {year} {1958})}\BibitemShut {NoStop}%
\bibitem [{\citenamefont {Mott}(1968)}]{mott68}%
  \BibitemOpen
  \bibfield  {author} {\bibinfo {author} {\bibfnamefont {N.~F.}\ \bibnamefont
  {Mott}},\ }\bibfield  {title} {\enquote {\bibinfo {title} {Conduction in
  non-crystalline systems},}\ }\href {\doibase 10.1080/14786436808223200}
  {\bibfield  {journal} {\bibinfo  {journal} {Phil. Mag.}\ }\textbf {\bibinfo
  {volume} {17}},\ \bibinfo {pages} {1259} (\bibinfo {year}
  {1968})}\BibitemShut {NoStop}%
\bibitem [{\citenamefont {Fleishman}\ and\ \citenamefont
  {Anderson}(1980)}]{fleishman80}%
  \BibitemOpen
  \bibfield  {author} {\bibinfo {author} {\bibfnamefont {L.}~\bibnamefont
  {Fleishman}}\ and\ \bibinfo {author} {\bibfnamefont {P.~W.}\ \bibnamefont
  {Anderson}},\ }\bibfield  {title} {\enquote {\bibinfo {title} {Interactions
  and the anderson transition},}\ }\href {\doibase 10.1103/PhysRevB.21.2366}
  {\bibfield  {journal} {\bibinfo  {journal} {Phys. Rev. B}\ }\textbf {\bibinfo
  {volume} {21}},\ \bibinfo {pages} {2366} (\bibinfo {year}
  {1980})}\BibitemShut {NoStop}%
\bibitem [{\citenamefont {Basko}\ \emph {et~al.}(2006)\citenamefont {Basko},
  \citenamefont {Aleiner},\ and\ \citenamefont {Altshuler}}]{basko06}%
  \BibitemOpen
  \bibfield  {author} {\bibinfo {author} {\bibfnamefont {D.~M.}\ \bibnamefont
  {Basko}}, \bibinfo {author} {\bibfnamefont {I.~L.}\ \bibnamefont {Aleiner}},
  \ and\ \bibinfo {author} {\bibfnamefont {B.~L.}\ \bibnamefont {Altshuler}},\
  }\bibfield  {title} {\enquote {\bibinfo {title} {Metal--insulator transition
  in a weakly interacting many-electron system with localized single-particle
  states},}\ }\href {\doibase 10.1016/j.aop.2005.11.014} {\bibfield  {journal}
  {\bibinfo  {journal} {Ann. Phys.}\ }\textbf {\bibinfo {volume} {321}},\
  \bibinfo {pages} {1126} (\bibinfo {year} {2006})}\BibitemShut {NoStop}%
\bibitem [{\citenamefont {Oganesyan}\ and\ \citenamefont
  {Huse}(2007)}]{oganesyan07}%
  \BibitemOpen
  \bibfield  {author} {\bibinfo {author} {\bibfnamefont {V.}~\bibnamefont
  {Oganesyan}}\ and\ \bibinfo {author} {\bibfnamefont {D.~A.}\ \bibnamefont
  {Huse}},\ }\bibfield  {title} {\enquote {\bibinfo {title} {Localization of
  interacting fermions at high temperature},}\ }\href {\doibase
  10.1103/PhysRevB.75.155111} {\bibfield  {journal} {\bibinfo  {journal} {Phys.
  Rev. B}\ }\textbf {\bibinfo {volume} {75}},\ \bibinfo {pages} {155111}
  (\bibinfo {year} {2007})}\BibitemShut {NoStop}%
\bibitem [{\citenamefont {Berkelbach}\ and\ \citenamefont
  {Reichman}(2010)}]{berkelbach10}%
  \BibitemOpen
  \bibfield  {author} {\bibinfo {author} {\bibfnamefont {T.~C.}\ \bibnamefont
  {Berkelbach}}\ and\ \bibinfo {author} {\bibfnamefont {D.~R.}\ \bibnamefont
  {Reichman}},\ }\bibfield  {title} {\enquote {\bibinfo {title} {Conductivity
  of disordered quantum lattice models at infinite temperature: Many-body
  localization},}\ }\href {\doibase 10.1103/PhysRevB.81.224429} {\bibfield
  {journal} {\bibinfo  {journal} {Phys. Rev. B}\ }\textbf {\bibinfo {volume}
  {81}},\ \bibinfo {pages} {224429} (\bibinfo {year} {2010})}\BibitemShut
  {NoStop}%
\bibitem [{\citenamefont {Bari\v{s}i\'{c}}\ and\ \citenamefont
  {Prelov\v{s}ek}(2010)}]{barisic10}%
  \BibitemOpen
  \bibfield  {author} {\bibinfo {author} {\bibfnamefont {O.~S.}\ \bibnamefont
  {Bari\v{s}i\'{c}}}\ and\ \bibinfo {author} {\bibfnamefont {P.}~\bibnamefont
  {Prelov\v{s}ek}},\ }\bibfield  {title} {\enquote {\bibinfo {title}
  {Conductivity in a disordered one-dimensional system of interacting
  fermions},}\ }\href {\doibase 10.1103/PhysRevB.82.161106} {\bibfield
  {journal} {\bibinfo  {journal} {Phys. Rev. B}\ }\textbf {\bibinfo {volume}
  {82}},\ \bibinfo {pages} {161106} (\bibinfo {year} {2010})}\BibitemShut
  {NoStop}%
\bibitem [{\citenamefont {Agarwal}\ \emph {et~al.}(2015)\citenamefont
  {Agarwal}, \citenamefont {Gopalakrishnan}, \citenamefont {Knap},
  \citenamefont {M\"{u}ller},\ and\ \citenamefont {Demler}}]{agarwal15}%
  \BibitemOpen
  \bibfield  {author} {\bibinfo {author} {\bibfnamefont {K.}~\bibnamefont
  {Agarwal}}, \bibinfo {author} {\bibfnamefont {S.}~\bibnamefont
  {Gopalakrishnan}}, \bibinfo {author} {\bibfnamefont {M.}~\bibnamefont
  {Knap}}, \bibinfo {author} {\bibfnamefont {M.}~\bibnamefont {M\"{u}ller}}, \
  and\ \bibinfo {author} {\bibfnamefont {E.}~\bibnamefont {Demler}},\
  }\bibfield  {title} {\enquote {\bibinfo {title} {Anomalous diffusion and
  griffiths effects near the many-body localization transition},}\ }\href
  {\doibase 10.1103/PhysRevLett.114.160401} {\bibfield  {journal} {\bibinfo
  {journal} {Phys. Rev. Lett.}\ }\textbf {\bibinfo {volume} {114}},\ \bibinfo
  {pages} {160401} (\bibinfo {year} {2015})}\BibitemShut {NoStop}%
\bibitem [{\citenamefont {Gopalakrishnan}\ \emph {et~al.}(2015)\citenamefont
  {Gopalakrishnan}, \citenamefont {M\"{u}ller}, \citenamefont {Khemani},
  \citenamefont {Knap}, \citenamefont {Demler},\ and\ \citenamefont
  {Huse}}]{gopal15}%
  \BibitemOpen
  \bibfield  {author} {\bibinfo {author} {\bibfnamefont {S.}~\bibnamefont
  {Gopalakrishnan}}, \bibinfo {author} {\bibfnamefont {M.}~\bibnamefont
  {M\"{u}ller}}, \bibinfo {author} {\bibfnamefont {V.}~\bibnamefont {Khemani}},
  \bibinfo {author} {\bibfnamefont {M.}~\bibnamefont {Knap}}, \bibinfo {author}
  {\bibfnamefont {E.}~\bibnamefont {Demler}}, \ and\ \bibinfo {author}
  {\bibfnamefont {D.~A.}\ \bibnamefont {Huse}},\ }\bibfield  {title} {\enquote
  {\bibinfo {title} {Low-frequency conductivity in many-body localized
  systems},}\ }\href {\doibase 10.1103/PhysRevB.92.104202} {\bibfield
  {journal} {\bibinfo  {journal} {Phys. Rev. B}\ }\textbf {\bibinfo {volume}
  {92}},\ \bibinfo {pages} {104202} (\bibinfo {year} {2015})}\BibitemShut
  {NoStop}%
\bibitem [{\citenamefont {Bar~Lev}\ \emph {et~al.}(2015)\citenamefont
  {Bar~Lev}, \citenamefont {Cohen},\ and\ \citenamefont {Reichman}}]{lev15}%
  \BibitemOpen
  \bibfield  {author} {\bibinfo {author} {\bibfnamefont {Y.}~\bibnamefont
  {Bar~Lev}}, \bibinfo {author} {\bibfnamefont {G.}~\bibnamefont {Cohen}}, \
  and\ \bibinfo {author} {\bibfnamefont {D.~R.}\ \bibnamefont {Reichman}},\
  }\bibfield  {title} {\enquote {\bibinfo {title} {Absence of diffusion in an
  interacting system of spinless fermions on a one-dimensional disordered
  lattice},}\ }\href {\doibase 10.1103/PhysRevLett.114.100601} {\bibfield
  {journal} {\bibinfo  {journal} {Phys. Rev. Lett.}\ }\textbf {\bibinfo
  {volume} {114}},\ \bibinfo {pages} {100601} (\bibinfo {year}
  {2015})}\BibitemShut {NoStop}%
\bibitem [{\citenamefont {Steinigeweg}\ \emph {et~al.}(2015)\citenamefont
  {Steinigeweg}, \citenamefont {Herbrych}, \citenamefont {Pollmann},\ and\
  \citenamefont {Brenig}}]{steinigeweg15}%
  \BibitemOpen
  \bibfield  {author} {\bibinfo {author} {\bibfnamefont {R.}~\bibnamefont
  {Steinigeweg}}, \bibinfo {author} {\bibfnamefont {J.}~\bibnamefont
  {Herbrych}}, \bibinfo {author} {\bibfnamefont {F.}~\bibnamefont {Pollmann}},
  \ and\ \bibinfo {author} {\bibfnamefont {W.}~\bibnamefont {Brenig}},\
  }\bibfield  {title} {\enquote {\bibinfo {title} {Scaling of the optical
  conductivity in the transition from thermal to many-body localized phases},}\
  }\href {http://arxiv.org/abs/1512.08519} {\bibfield  {journal} {\bibinfo
  {journal} {ArXiv e-prints}\ } (\bibinfo {year} {2015})},\ \Eprint
  {http://arxiv.org/abs/1512.08519} {1512.08519 [cond-mat.stat-mech]}
  \BibitemShut {NoStop}%
\bibitem [{\citenamefont {Bari\v{s}i\'{c}}\ \emph {et~al.}(2016)\citenamefont
  {Bari\v{s}i\'{c}}, \citenamefont {Kokalj}, \citenamefont {Balog},\ and\
  \citenamefont {Prelov\v{s}ek}}]{barisic16}%
  \BibitemOpen
  \bibfield  {author} {\bibinfo {author} {\bibfnamefont {O.~S.}\ \bibnamefont
  {Bari\v{s}i\'{c}}}, \bibinfo {author} {\bibfnamefont {J.}~\bibnamefont
  {Kokalj}}, \bibinfo {author} {\bibfnamefont {I.}~\bibnamefont {Balog}}, \
  and\ \bibinfo {author} {\bibfnamefont {P.}~\bibnamefont {Prelov\v{s}ek}},\
  }\bibfield  {title} {\enquote {\bibinfo {title} {Dynamical conductivity and
  its fluctuations along the crossover to many-body localization},}\ }\href
  {\doibase 10.1103/PhysRevB.94.045126} {\bibfield  {journal} {\bibinfo
  {journal} {Phys. Rev. B}\ }\textbf {\bibinfo {volume} {94}},\ \bibinfo
  {pages} {045126} (\bibinfo {year} {2016})}\BibitemShut {NoStop}%
\bibitem [{\citenamefont {Pal}\ and\ \citenamefont {Huse}(2010)}]{pal10}%
  \BibitemOpen
  \bibfield  {author} {\bibinfo {author} {\bibfnamefont {A.}~\bibnamefont
  {Pal}}\ and\ \bibinfo {author} {\bibfnamefont {D.~A.}\ \bibnamefont {Huse}},\
  }\bibfield  {title} {\enquote {\bibinfo {title} {Many-body localization phase
  transition},}\ }\href {\doibase 10.1103/PhysRevB.82.174411} {\bibfield
  {journal} {\bibinfo  {journal} {Phys. Rev. B}\ }\textbf {\bibinfo {volume}
  {82}},\ \bibinfo {pages} {174411} (\bibinfo {year} {2010})}\BibitemShut
  {NoStop}%
\bibitem [{\citenamefont {Serbyn}\ \emph {et~al.}(2013)\citenamefont {Serbyn},
  \citenamefont {Papi\'{c}},\ and\ \citenamefont {Abanin}}]{serbyn131}%
  \BibitemOpen
  \bibfield  {author} {\bibinfo {author} {\bibfnamefont {M.}~\bibnamefont
  {Serbyn}}, \bibinfo {author} {\bibfnamefont {Z.}~\bibnamefont {Papi\'{c}}}, \
  and\ \bibinfo {author} {\bibfnamefont {D.~A.}\ \bibnamefont {Abanin}},\
  }\bibfield  {title} {\enquote {\bibinfo {title} {Local conservation laws and
  the structure of the many-body localized states},}\ }\href {\doibase
  10.1103/PhysRevLett.111.127201} {\bibfield  {journal} {\bibinfo  {journal}
  {Phys. Rev. Lett.}\ }\textbf {\bibinfo {volume} {111}},\ \bibinfo {pages}
  {127201} (\bibinfo {year} {2013})}\BibitemShut {NoStop}%
\bibitem [{\citenamefont {De~Luca}\ and\ \citenamefont
  {Scardicchio}(2013)}]{deluca13}%
  \BibitemOpen
  \bibfield  {author} {\bibinfo {author} {\bibfnamefont {A.}~\bibnamefont
  {De~Luca}}\ and\ \bibinfo {author} {\bibfnamefont {A.}~\bibnamefont
  {Scardicchio}},\ }\bibfield  {title} {\enquote {\bibinfo {title} {Ergodicity
  breaking in a model showing many-body localization},}\ }\href
  {http://stacks.iop.org/0295-5075/101/i=3/a=37003} {\bibfield  {journal}
  {\bibinfo  {journal} {EPL (Europhysics Letters)}\ }\textbf {\bibinfo {volume}
  {101}},\ \bibinfo {pages} {37003} (\bibinfo {year} {2013})}\BibitemShut
  {NoStop}%
\bibitem [{\citenamefont {Huse}\ \emph {et~al.}(2014)\citenamefont {Huse},
  \citenamefont {Nandkishore},\ and\ \citenamefont {Oganesyan}}]{huse14}%
  \BibitemOpen
  \bibfield  {author} {\bibinfo {author} {\bibfnamefont {D.~A.}\ \bibnamefont
  {Huse}}, \bibinfo {author} {\bibfnamefont {R.}~\bibnamefont {Nandkishore}}, \
  and\ \bibinfo {author} {\bibfnamefont {V.}~\bibnamefont {Oganesyan}},\
  }\bibfield  {title} {\enquote {\bibinfo {title} {Phenomenology of fully
  many-body-localized systems},}\ }\href {\doibase 10.1103/PhysRevB.90.174202}
  {\bibfield  {journal} {\bibinfo  {journal} {Phys. Rev. B}\ }\textbf {\bibinfo
  {volume} {90}},\ \bibinfo {pages} {174202} (\bibinfo {year}
  {2014})}\BibitemShut {NoStop}%
\bibitem [{\citenamefont {Vosk}\ and\ \citenamefont {Altman}(2013)}]{vosk13}%
  \BibitemOpen
  \bibfield  {author} {\bibinfo {author} {\bibfnamefont {R.}~\bibnamefont
  {Vosk}}\ and\ \bibinfo {author} {\bibfnamefont {E.}~\bibnamefont {Altman}},\
  }\bibfield  {title} {\enquote {\bibinfo {title} {Many-body localization in
  one dimension as a dynamical renormalization group fixed point},}\ }\href
  {\doibase 10.1103/PhysRevLett.110.067204} {\bibfield  {journal} {\bibinfo
  {journal} {Phys. Rev. Lett.}\ }\textbf {\bibinfo {volume} {110}},\ \bibinfo
  {pages} {067204} (\bibinfo {year} {2013})}\BibitemShut {NoStop}%
\bibitem [{\citenamefont {Huse}\ \emph {et~al.}(2013)\citenamefont {Huse},
  \citenamefont {Nandkishore}, \citenamefont {Oganesyan}, \citenamefont {Pal},\
  and\ \citenamefont {Sondhi}}]{vosk14}%
  \BibitemOpen
  \bibfield  {author} {\bibinfo {author} {\bibfnamefont {D.~A.}\ \bibnamefont
  {Huse}}, \bibinfo {author} {\bibfnamefont {R.}~\bibnamefont {Nandkishore}},
  \bibinfo {author} {\bibfnamefont {V.}~\bibnamefont {Oganesyan}}, \bibinfo
  {author} {\bibfnamefont {A.}~\bibnamefont {Pal}}, \ and\ \bibinfo {author}
  {\bibfnamefont {S.~L.}\ \bibnamefont {Sondhi}},\ }\bibfield  {title}
  {\enquote {\bibinfo {title} {Localization-protected quantum order},}\ }\href
  {\doibase 10.1103/PhysRevB.88.014206} {\bibfield  {journal} {\bibinfo
  {journal} {Phys. Rev. B}\ }\textbf {\bibinfo {volume} {88}},\ \bibinfo
  {pages} {014206} (\bibinfo {year} {2013})}\BibitemShut {NoStop}%
\bibitem [{\citenamefont {Serbyn}\ \emph {et~al.}(2014)\citenamefont {Serbyn},
  \citenamefont {Papi\'{c}},\ and\ \citenamefont {Abanin}}]{serbyn14}%
  \BibitemOpen
  \bibfield  {author} {\bibinfo {author} {\bibfnamefont {M.}~\bibnamefont
  {Serbyn}}, \bibinfo {author} {\bibfnamefont {Z.}~\bibnamefont {Papi\'{c}}}, \
  and\ \bibinfo {author} {\bibfnamefont {D.~A.}\ \bibnamefont {Abanin}},\
  }\bibfield  {title} {\enquote {\bibinfo {title} {Quantum quenches in the
  many-body localized phase},}\ }\href {\doibase 10.1103/PhysRevB.90.174302}
  {\bibfield  {journal} {\bibinfo  {journal} {Phys. Rev. B}\ }\textbf {\bibinfo
  {volume} {90}},\ \bibinfo {pages} {174302} (\bibinfo {year}
  {2014})}\BibitemShut {NoStop}%
\bibitem [{\citenamefont {Vasseur}\ \emph {et~al.}(2015)\citenamefont
  {Vasseur}, \citenamefont {Parameswaran},\ and\ \citenamefont
  {Moore}}]{vasseur15}%
  \BibitemOpen
  \bibfield  {author} {\bibinfo {author} {\bibfnamefont {R.}~\bibnamefont
  {Vasseur}}, \bibinfo {author} {\bibfnamefont {S.~A.}\ \bibnamefont
  {Parameswaran}}, \ and\ \bibinfo {author} {\bibfnamefont {J.~E.}\
  \bibnamefont {Moore}},\ }\bibfield  {title} {\enquote {\bibinfo {title}
  {Quantum revivals and many-body localization},}\ }\href {\doibase
  10.1103/PhysRevB.91.140202} {\bibfield  {journal} {\bibinfo  {journal} {Phys.
  Rev. B}\ }\textbf {\bibinfo {volume} {91}},\ \bibinfo {pages} {140202}
  (\bibinfo {year} {2015})}\BibitemShut {NoStop}%
\bibitem [{\citenamefont {Kondov}\ \emph {et~al.}(2015)\citenamefont {Kondov},
  \citenamefont {McGehee}, \citenamefont {Xu},\ and\ \citenamefont
  {DeMarco}}]{kondov15}%
  \BibitemOpen
  \bibfield  {author} {\bibinfo {author} {\bibfnamefont {S.~S.}\ \bibnamefont
  {Kondov}}, \bibinfo {author} {\bibfnamefont {W.~R.}\ \bibnamefont {McGehee}},
  \bibinfo {author} {\bibfnamefont {W.}~\bibnamefont {Xu}}, \ and\ \bibinfo
  {author} {\bibfnamefont {B.}~\bibnamefont {DeMarco}},\ }\bibfield  {title}
  {\enquote {\bibinfo {title} {Disorder-induced localization in a strongly
  correlated atomic hubbard gas},}\ }\href {\doibase
  10.1103/PhysRevLett.114.083002} {\bibfield  {journal} {\bibinfo  {journal}
  {Phys. Rev. Lett.}\ }\textbf {\bibinfo {volume} {114}},\ \bibinfo {pages}
  {083002} (\bibinfo {year} {2015})}\BibitemShut {NoStop}%
\bibitem [{\citenamefont {Schreiber}\ \emph {et~al.}(2015)\citenamefont
  {Schreiber}, \citenamefont {Hodgman}, \citenamefont {Bordia}, \citenamefont
  {L\"{u}schen}, \citenamefont {Fischer}, \citenamefont {Vosk}, \citenamefont
  {Altman}, \citenamefont {Schneider},\ and\ \citenamefont
  {Bloch}}]{schreiber15}%
  \BibitemOpen
  \bibfield  {author} {\bibinfo {author} {\bibfnamefont {M.}~\bibnamefont
  {Schreiber}}, \bibinfo {author} {\bibfnamefont {S.~S.}\ \bibnamefont
  {Hodgman}}, \bibinfo {author} {\bibfnamefont {P.}~\bibnamefont {Bordia}},
  \bibinfo {author} {\bibfnamefont {H.~P.}\ \bibnamefont {L\"{u}schen}},
  \bibinfo {author} {\bibfnamefont {M.~H.}\ \bibnamefont {Fischer}}, \bibinfo
  {author} {\bibfnamefont {R.}~\bibnamefont {Vosk}}, \bibinfo {author}
  {\bibfnamefont {E.}~\bibnamefont {Altman}}, \bibinfo {author} {\bibfnamefont
  {U.}~\bibnamefont {Schneider}}, \ and\ \bibinfo {author} {\bibfnamefont
  {I.}~\bibnamefont {Bloch}},\ }\bibfield  {title} {\enquote {\bibinfo {title}
  {Observation of many-body localization of interacting fermions in a
  quasi-random optical lattice},}\ }\href {\doibase 10.1126/science.aaa7432}
  {\bibfield  {journal} {\bibinfo  {journal} {Science}\ }\textbf {\bibinfo
  {volume} {349}},\ \bibinfo {pages} {842} (\bibinfo {year}
  {2015})}\BibitemShut {NoStop}%
\bibitem [{\citenamefont {Bordia}\ \emph {et~al.}(2016)\citenamefont {Bordia},
  \citenamefont {L\"{u}schen}, \citenamefont {Hodgman}, \citenamefont
  {Schreiber}, \citenamefont {Bloch},\ and\ \citenamefont
  {Schneider}}]{bordia16}%
  \BibitemOpen
  \bibfield  {author} {\bibinfo {author} {\bibfnamefont {P.}~\bibnamefont
  {Bordia}}, \bibinfo {author} {\bibfnamefont {H.~P.}\ \bibnamefont
  {L\"{u}schen}}, \bibinfo {author} {\bibfnamefont {S.~S.}\ \bibnamefont
  {Hodgman}}, \bibinfo {author} {\bibfnamefont {M.}~\bibnamefont {Schreiber}},
  \bibinfo {author} {\bibfnamefont {I.}~\bibnamefont {Bloch}}, \ and\ \bibinfo
  {author} {\bibfnamefont {U.}~\bibnamefont {Schneider}},\ }\bibfield  {title}
  {\enquote {\bibinfo {title} {Coupling identical 1$\mathrm{D}$ many-body
  localized systems},}\ }\href {\doibase 10.1103/PhysRevLett.116.140401}
  {\bibfield  {journal} {\bibinfo  {journal} {Phys. Rev. Lett.}\ }\textbf
  {\bibinfo {volume} {116}},\ \bibinfo {pages} {140401} (\bibinfo {year}
  {2016})}\BibitemShut {NoStop}%
\bibitem [{\citenamefont {L{\"{u}}schen}\ \emph {et~al.}(2016)\citenamefont
  {L{\"{u}}schen}, \citenamefont {Bordia}, \citenamefont {Scherg},
  \citenamefont {Alet}, \citenamefont {Altman}, \citenamefont {Schneider},\
  and\ \citenamefont {Bloch}}]{luschen16}%
  \BibitemOpen
  \bibfield  {author} {\bibinfo {author} {\bibfnamefont {H.~P.}\ \bibnamefont
  {L{\"{u}}schen}}, \bibinfo {author} {\bibfnamefont {P.}~\bibnamefont
  {Bordia}}, \bibinfo {author} {\bibfnamefont {S.}~\bibnamefont {Scherg}},
  \bibinfo {author} {\bibfnamefont {F.}~\bibnamefont {Alet}}, \bibinfo {author}
  {\bibfnamefont {E.}~\bibnamefont {Altman}}, \bibinfo {author} {\bibfnamefont
  {U.}~\bibnamefont {Schneider}}, \ and\ \bibinfo {author} {\bibfnamefont
  {I.}~\bibnamefont {Bloch}},\ }\bibfield  {title} {\enquote {\bibinfo {title}
  {Evidence for griffiths-type dynamics near the many-body localization
  transition in quasi-periodic systems},}\ }\href
  {http://arxiv.org/abs/1612.07173} {\bibfield  {journal} {\bibinfo  {journal}
  {ArXiv e-prints}\ } (\bibinfo {year} {2016})},\ \Eprint
  {http://arxiv.org/abs/1612.07173} {1612.07173 [cond-mat.stat-mech]}
  \BibitemShut {NoStop}%
\bibitem [{\citenamefont {Luitz}\ \emph {et~al.}(2016)\citenamefont {Luitz},
  \citenamefont {Laflorencie},\ and\ \citenamefont {Alet}}]{luitz16}%
  \BibitemOpen
  \bibfield  {author} {\bibinfo {author} {\bibfnamefont {D.~J.}\ \bibnamefont
  {Luitz}}, \bibinfo {author} {\bibfnamefont {N.}~\bibnamefont {Laflorencie}},
  \ and\ \bibinfo {author} {\bibfnamefont {F.}~\bibnamefont {Alet}},\
  }\bibfield  {title} {\enquote {\bibinfo {title} {Extended slow dynamical
  regime prefiguring the many-body localization transition},}\ }\href {\doibase
  10.1103/PhysRevB.93.060201} {\bibfield  {journal} {\bibinfo  {journal} {Phys.
  Rev. B}\ }\textbf {\bibinfo {volume} {93}},\ \bibinfo {pages} {060201}
  (\bibinfo {year} {2016})}\BibitemShut {NoStop}%
\bibitem [{\citenamefont {\v{Z}nidari\v{c}}\ \emph {et~al.}(2016)\citenamefont
  {\v{Z}nidari\v{c}}, \citenamefont {Scardicchio},\ and\ \citenamefont
  {Varma}}]{znidaric16}%
  \BibitemOpen
  \bibfield  {author} {\bibinfo {author} {\bibfnamefont {M.}~\bibnamefont
  {\v{Z}nidari\v{c}}}, \bibinfo {author} {\bibfnamefont {A.}~\bibnamefont
  {Scardicchio}}, \ and\ \bibinfo {author} {\bibfnamefont {V.~K.}\ \bibnamefont
  {Varma}},\ }\bibfield  {title} {\enquote {\bibinfo {title} {Diffusive and
  subdiffusive spin transport in the ergodic phase of a many-body localizable
  system},}\ }\href {\doibase 10.1103/PhysRevLett.117.040601} {\bibfield
  {journal} {\bibinfo  {journal} {Phys. Rev. Lett.}\ }\textbf {\bibinfo
  {volume} {117}},\ \bibinfo {pages} {040601} (\bibinfo {year}
  {2016})}\BibitemShut {NoStop}%
\bibitem [{\citenamefont {Luitz}\ and\ \citenamefont {Lev}(2016)}]{luitz116}%
  \BibitemOpen
  \bibfield  {author} {\bibinfo {author} {\bibfnamefont {D.~J.}\ \bibnamefont
  {Luitz}}\ and\ \bibinfo {author} {\bibfnamefont {Y.~B.}\ \bibnamefont
  {Lev}},\ }\bibfield  {title} {\enquote {\bibinfo {title} {The ergodic side of
  the many-body localization transition},}\ }\href
  {http://arxiv.org/abs/1610.08993} {\bibfield  {journal} {\bibinfo  {journal}
  {ArXiv e-prints}\ } (\bibinfo {year} {2016})},\ \Eprint
  {http://arxiv.org/abs/1610.08993} {1610.08993 [cond-mat.stat-mech]}
  \BibitemShut {NoStop}%
\bibitem [{\citenamefont {Vosk}\ \emph {et~al.}(2015)\citenamefont {Vosk},
  \citenamefont {Huse},\ and\ \citenamefont {Altman}}]{vosk15}%
  \BibitemOpen
  \bibfield  {author} {\bibinfo {author} {\bibfnamefont {R.}~\bibnamefont
  {Vosk}}, \bibinfo {author} {\bibfnamefont {D.~A.}\ \bibnamefont {Huse}}, \
  and\ \bibinfo {author} {\bibfnamefont {E.}~\bibnamefont {Altman}},\
  }\bibfield  {title} {\enquote {\bibinfo {title} {Theory of the many-body
  localization transition in one-dimensional systems},}\ }\href {\doibase
  10.1103/PhysRevX.5.031032} {\bibfield  {journal} {\bibinfo  {journal} {Phys.
  Rev. X}\ }\textbf {\bibinfo {volume} {5}},\ \bibinfo {pages} {031032}
  (\bibinfo {year} {2015})}\BibitemShut {NoStop}%
\bibitem [{\citenamefont {Luitz}\ \emph {et~al.}(2015)\citenamefont {Luitz},
  \citenamefont {Laflorencie},\ and\ \citenamefont {Alet}}]{luitz15}%
  \BibitemOpen
  \bibfield  {author} {\bibinfo {author} {\bibfnamefont {D.~J.}\ \bibnamefont
  {Luitz}}, \bibinfo {author} {\bibfnamefont {N.}~\bibnamefont {Laflorencie}},
  \ and\ \bibinfo {author} {\bibfnamefont {F.}~\bibnamefont {Alet}},\
  }\bibfield  {title} {\enquote {\bibinfo {title} {Many-body localization edge
  in the random-field heisenberg chain},}\ }\href {\doibase
  10.1103/PhysRevB.91.081103} {\bibfield  {journal} {\bibinfo  {journal} {Phys.
  Rev. B}\ }\textbf {\bibinfo {volume} {91}},\ \bibinfo {pages} {081103}
  (\bibinfo {year} {2015})}\BibitemShut {NoStop}%
\bibitem [{\citenamefont {Mierzejewski}\ \emph {et~al.}(2016)\citenamefont
  {Mierzejewski}, \citenamefont {Herbrych},\ and\ \citenamefont
  {Prelov\v{s}ek}}]{mierzejewski16}%
  \BibitemOpen
  \bibfield  {author} {\bibinfo {author} {\bibfnamefont {M.}~\bibnamefont
  {Mierzejewski}}, \bibinfo {author} {\bibfnamefont {J.}~\bibnamefont
  {Herbrych}}, \ and\ \bibinfo {author} {\bibfnamefont {P.}~\bibnamefont
  {Prelov\v{s}ek}},\ }\bibfield  {title} {\enquote {\bibinfo {title} {Universal
  dynamics of density correlations at the transition to many-body localized
  state},}\ }\href {http://arxiv.org/abs/1607.04992} {\bibfield  {journal}
  {\bibinfo  {journal} {ArXiv e-prints}\ } (\bibinfo {year} {2016})},\ \Eprint
  {http://arxiv.org/abs/1607.04992} {1607.04992 [cond-mat.stat-mech]}
  \BibitemShut {NoStop}%
\bibitem [{\citenamefont {Vollhardt}\ and\ \citenamefont
  {W\"{o}lfle}(1980)}]{vollhardt80}%
  \BibitemOpen
  \bibfield  {author} {\bibinfo {author} {\bibfnamefont {D.}~\bibnamefont
  {Vollhardt}}\ and\ \bibinfo {author} {\bibfnamefont {P.}~\bibnamefont
  {W\"{o}lfle}},\ }\bibfield  {title} {\enquote {\bibinfo {title} {Anderson
  localization in d<~2 dimensions: A self-consistent diagrammatic theory},}\
  }\href {\doibase 10.1103/PhysRevLett.45.842} {\bibfield  {journal} {\bibinfo
  {journal} {Phys. Rev. Lett.}\ }\textbf {\bibinfo {volume} {45}},\ \bibinfo
  {pages} {842} (\bibinfo {year} {1980})}\BibitemShut {NoStop}%
\bibitem [{\citenamefont {D.}\ and\ \citenamefont
  {W\"{o}lfle}(1980)}]{vollhardt801}%
  \BibitemOpen
  \bibfield  {author} {\bibinfo {author} {\bibfnamefont {Vollhardt}\
  \bibnamefont {D.}}\ and\ \bibinfo {author} {\bibfnamefont {P.}~\bibnamefont
  {W\"{o}lfle}},\ }\bibfield  {title} {\enquote {\bibinfo {title}
  {Diagrammatic, self-consistent treatment of the anderson localization problem
  in d≤2 dimensions},}\ }\href {\doibase 10.1103/PhysRevB.22.4666} {\bibfield
   {journal} {\bibinfo  {journal} {Phys. Rev. B}\ }\textbf {\bibinfo {volume}
  {22}},\ \bibinfo {pages} {4666} (\bibinfo {year} {1980})}\BibitemShut
  {NoStop}%
\bibitem [{\citenamefont {Gopalakrishnan}\ \emph {et~al.}(2016)\citenamefont
  {Gopalakrishnan}, \citenamefont {Agarwal}, \citenamefont {Demler},
  \citenamefont {Huse},\ and\ \citenamefont {Knap}}]{gopal16}%
  \BibitemOpen
  \bibfield  {author} {\bibinfo {author} {\bibfnamefont {S.}~\bibnamefont
  {Gopalakrishnan}}, \bibinfo {author} {\bibfnamefont {K.}~\bibnamefont
  {Agarwal}}, \bibinfo {author} {\bibfnamefont {E.~A.}\ \bibnamefont {Demler}},
  \bibinfo {author} {\bibfnamefont {D.~A.}\ \bibnamefont {Huse}}, \ and\
  \bibinfo {author} {\bibfnamefont {M.}~\bibnamefont {Knap}},\ }\bibfield
  {title} {\enquote {\bibinfo {title} {Griffiths effects and slow dynamics in
  nearly many-body localized systems},}\ }\href {\doibase
  10.1103/PhysRevB.93.134206} {\bibfield  {journal} {\bibinfo  {journal} {Phys.
  Rev. B}\ }\textbf {\bibinfo {volume} {93}},\ \bibinfo {pages} {134206}
  (\bibinfo {year} {2016})}\BibitemShut {NoStop}%
\bibitem [{\citenamefont {Pirc}\ and\ \citenamefont {Dick}(1974)}]{pirc74}%
  \BibitemOpen
  \bibfield  {author} {\bibinfo {author} {\bibfnamefont {R.}~\bibnamefont
  {Pirc}}\ and\ \bibinfo {author} {\bibfnamefont {G.}~\bibnamefont {Dick}},\
  }\bibfield  {title} {\enquote {\bibinfo {title} {Exact isolated and
  isothermal susceptibilities for an interacting dipole-lattice system},}\
  }\href {\doibase 10.1103/PhysRevB.9.2701} {\bibfield  {journal} {\bibinfo
  {journal} {Phys. Rev. B}\ }\textbf {\bibinfo {volume} {9}},\ \bibinfo {pages}
  {2701} (\bibinfo {year} {1974})}\BibitemShut {NoStop}%
\bibitem [{\citenamefont {G\"{o}tze}(1979)}]{gotze79}%
  \BibitemOpen
  \bibfield  {author} {\bibinfo {author} {\bibfnamefont {W.}~\bibnamefont
  {G\"{o}tze}},\ }\bibfield  {title} {\enquote {\bibinfo {title} {Strong
  three-dimensional random potential},}\ }\href {\doibase
  10.1088/0022-3719/12/7/018} {\bibfield  {journal} {\bibinfo  {journal} {J.
  Phys. C: Solid State Phys.}\ }\textbf {\bibinfo {volume} {12}},\ \bibinfo
  {pages} {1279} (\bibinfo {year} {1979})}\BibitemShut {NoStop}%
\bibitem [{\citenamefont {Forster}(1995)}]{forster95}%
  \BibitemOpen
  \bibfield  {author} {\bibinfo {author} {\bibfnamefont {D.}~\bibnamefont
  {Forster}},\ }\href@noop {} {\emph {\bibinfo {title} {Hydrodynamic
  Fluctuations, Broken Symmetry, And Correlation Functions}}}\ (\bibinfo
  {publisher} {Westview Press, New York},\ \bibinfo {year} {1995})\BibitemShut
  {NoStop}%
\bibitem [{\citenamefont {Khait}\ \emph {et~al.}(2016)\citenamefont {Khait},
  \citenamefont {Gazit}, \citenamefont {Yao},\ and\ \citenamefont
  {Auerbach}}]{khait16}%
  \BibitemOpen
  \bibfield  {author} {\bibinfo {author} {\bibfnamefont {I.}~\bibnamefont
  {Khait}}, \bibinfo {author} {\bibfnamefont {S.}~\bibnamefont {Gazit}},
  \bibinfo {author} {\bibfnamefont {N.~Y.}\ \bibnamefont {Yao}}, \ and\
  \bibinfo {author} {\bibfnamefont {A.}~\bibnamefont {Auerbach}},\ }\bibfield
  {title} {\enquote {\bibinfo {title} {Spin transport of weakly disordered
  heisenberg chain at infinite temperature},}\ }\href {\doibase
  10.1103/PhysRevB.93.224205} {\bibfield  {journal} {\bibinfo  {journal} {Phys.
  Rev. B}\ }\textbf {\bibinfo {volume} {93}},\ \bibinfo {pages} {224205}
  (\bibinfo {year} {2016})}\BibitemShut {NoStop}%
\bibitem [{\citenamefont {Long}\ \emph {et~al.}(2003)\citenamefont {Long},
  \citenamefont {Prelov\v{s}ek}, \citenamefont {El~Shawish}, \citenamefont
  {Karadamoglou},\ and\ \citenamefont {Zotos}}]{long03}%
  \BibitemOpen
  \bibfield  {author} {\bibinfo {author} {\bibfnamefont {M.~W.}\ \bibnamefont
  {Long}}, \bibinfo {author} {\bibfnamefont {P.}~\bibnamefont {Prelov\v{s}ek}},
  \bibinfo {author} {\bibfnamefont {S.}~\bibnamefont {El~Shawish}}, \bibinfo
  {author} {\bibfnamefont {J.}~\bibnamefont {Karadamoglou}}, \ and\ \bibinfo
  {author} {\bibfnamefont {X.}~\bibnamefont {Zotos}},\ }\bibfield  {title}
  {\enquote {\bibinfo {title} {Finite-temperature dynamical correlations using
  the microcanonical ensemble and the lanczos algorithm},}\ }\href {\doibase
  10.1103/PhysRevB.68.235106} {\bibfield  {journal} {\bibinfo  {journal} {Phys.
  Rev. B}\ }\textbf {\bibinfo {volume} {68}},\ \bibinfo {pages} {235106}
  (\bibinfo {year} {2003})}\BibitemShut {NoStop}%
\bibitem [{\citenamefont {Prelov\v{s}ek}\ and\ \citenamefont
  {Bon\v{c}a}(2013)}]{prelovsek13}%
  \BibitemOpen
  \bibfield  {author} {\bibinfo {author} {\bibfnamefont {P.}~\bibnamefont
  {Prelov\v{s}ek}}\ and\ \bibinfo {author} {\bibfnamefont {J.}~\bibnamefont
  {Bon\v{c}a}},\ }\bibfield  {title} {\enquote {\bibinfo {title} {Ground state
  and finite temperature lanczos methods},}\ }in\ \href@noop {} {\emph
  {\bibinfo {booktitle} {Strongly Correlated Systems - Numerical Methods}}},\
  \bibinfo {editor} {edited by\ \bibinfo {editor} {\bibfnamefont
  {A.}~\bibnamefont {Avella}}\ and\ \bibinfo {editor} {\bibfnamefont
  {F.}~\bibnamefont {Mancini}}}\ (\bibinfo  {publisher} {Springer},\ \bibinfo
  {address} {Berlin},\ \bibinfo {year} {2013})\BibitemShut {NoStop}%
\bibitem [{\citenamefont {Herbrych}\ \emph {et~al.}(2012)\citenamefont
  {Herbrych}, \citenamefont {Steinigeweg},\ and\ \citenamefont
  {Prelov\v{s}ek}}]{herbrych12}%
  \BibitemOpen
  \bibfield  {author} {\bibinfo {author} {\bibfnamefont {J.}~\bibnamefont
  {Herbrych}}, \bibinfo {author} {\bibfnamefont {R.}~\bibnamefont
  {Steinigeweg}}, \ and\ \bibinfo {author} {\bibfnamefont {P.}~\bibnamefont
  {Prelov\v{s}ek}},\ }\bibfield  {title} {\enquote {\bibinfo {title} {Spin
  hydrodynamics in the s=1/2 anisotropic heisenberg chain},}\ }\href {\doibase
  10.1103/PhysRevB.86.115106} {\bibfield  {journal} {\bibinfo  {journal} {Phys.
  Rev. B}\ }\textbf {\bibinfo {volume} {86}},\ \bibinfo {pages} {115106}
  (\bibinfo {year} {2012})}\BibitemShut {NoStop}%
\bibitem [{\citenamefont {G\"{o}tze}\ and\ \citenamefont
  {W\"{o}lfle}(1972)}]{gotze72}%
  \BibitemOpen
  \bibfield  {author} {\bibinfo {author} {\bibfnamefont {W.}~\bibnamefont
  {G\"{o}tze}}\ and\ \bibinfo {author} {\bibfnamefont {P.}~\bibnamefont
  {W\"{o}lfle}},\ }\bibfield  {title} {\enquote {\bibinfo {title} {Homogeneous
  dynamical conductivity of simple metals},}\ }\href {\doibase
  10.1103/PhysRevB.6.1226} {\bibfield  {journal} {\bibinfo  {journal} {Phys.
  Rev. B}\ }\textbf {\bibinfo {volume} {6}},\ \bibinfo {pages} {1226} (\bibinfo
  {year} {1972})}\BibitemShut {NoStop}%
\bibitem [{\citenamefont {Mori}(1965)}]{mori65}%
  \BibitemOpen
  \bibfield  {author} {\bibinfo {author} {\bibfnamefont {H.}~\bibnamefont
  {Mori}},\ }\bibfield  {title} {\enquote {\bibinfo {title} {Transport,
  collective motion, and brownian motion},}\ }\href {\doibase
  10.1143/PTP.33.423} {\bibfield  {journal} {\bibinfo  {journal} {Prog. Theor.
  Phys.}\ }\textbf {\bibinfo {volume} {33}},\ \bibinfo {pages} {423} (\bibinfo
  {year} {1965})}\BibitemShut {NoStop}%
\bibitem [{\citenamefont {Karahalios}\ \emph {et~al.}(2009)\citenamefont
  {Karahalios}, \citenamefont {Metavitsiadis}, \citenamefont {Zotos},
  \citenamefont {Gorczyca},\ and\ \citenamefont
  {Prelov\v{s}ek}}]{karahalios09}%
  \BibitemOpen
  \bibfield  {author} {\bibinfo {author} {\bibfnamefont {A.}~\bibnamefont
  {Karahalios}}, \bibinfo {author} {\bibfnamefont {A.}~\bibnamefont
  {Metavitsiadis}}, \bibinfo {author} {\bibfnamefont {X.}~\bibnamefont
  {Zotos}}, \bibinfo {author} {\bibfnamefont {A.}~\bibnamefont {Gorczyca}}, \
  and\ \bibinfo {author} {\bibfnamefont {P.}~\bibnamefont {Prelov\v{s}ek}},\
  }\bibfield  {title} {\enquote {\bibinfo {title} {Finite-temperature transport
  in disordered heisenberg chains},}\ }\href {\doibase
  10.1103/PhysRevB.79.024425} {\bibfield  {journal} {\bibinfo  {journal} {Phys.
  Rev. B}\ }\textbf {\bibinfo {volume} {79}},\ \bibinfo {pages} {024425}
  (\bibinfo {year} {2009})}\BibitemShut {NoStop}%
\bibitem [{\citenamefont {Prelov\v{s}ek}\ \emph {et~al.}(2016)\citenamefont
  {Prelov\v{s}ek}, \citenamefont {Mierzejewski}, \citenamefont
  {Bari{\v{s}}i{\'{c}}},\ and\ \citenamefont {Herbrych}}]{prelovsek17}%
  \BibitemOpen
  \bibfield  {author} {\bibinfo {author} {\bibfnamefont {P.}~\bibnamefont
  {Prelov\v{s}ek}}, \bibinfo {author} {\bibfnamefont {M.}~\bibnamefont
  {Mierzejewski}}, \bibinfo {author} {\bibfnamefont {O.}~\bibnamefont
  {Bari{\v{s}}i{\'{c}}}}, \ and\ \bibinfo {author} {\bibfnamefont
  {J.}~\bibnamefont {Herbrych}},\ }\bibfield  {title} {\enquote {\bibinfo
  {title} {Density correlations and transport in models of many-body
  localization},}\ }\href {http://arxiv.org/abs/1611.03611} {\bibfield
  {journal} {\bibinfo  {journal} {ArXiv e-prints}\ } (\bibinfo {year}
  {2016})},\ \Eprint {http://arxiv.org/abs/1611.03611} {1611.03611
  [cond-mat.stat-mech]} \BibitemShut {NoStop}%
\bibitem [{\citenamefont {Prelov{\v{s}}ek}\ \emph {et~al.}(2016)\citenamefont
  {Prelov{\v{s}}ek}, \citenamefont {Bari{\v{s}}i{\'{c}}},\ and\ \citenamefont
  {{\v{Z}}nidari{\v{c}}}}]{prelovsek16}%
  \BibitemOpen
  \bibfield  {author} {\bibinfo {author} {\bibfnamefont {P.}~\bibnamefont
  {Prelov{\v{s}}ek}}, \bibinfo {author} {\bibfnamefont {O.~S.}\ \bibnamefont
  {Bari{\v{s}}i{\'{c}}}}, \ and\ \bibinfo {author} {\bibfnamefont
  {M.}~\bibnamefont {{\v{Z}}nidari{\v{c}}}},\ }\bibfield  {title} {\enquote
  {\bibinfo {title} {Absence of full many-body localization in the disordered
  hubbard chain},}\ }\href {\doibase 10.1103/PhysRevB.94.241104} {\bibfield
  {journal} {\bibinfo  {journal} {Phys. Rev. B}\ }\textbf {\bibinfo {volume}
  {94}},\ \bibinfo {pages} {241104} (\bibinfo {year} {2016})}\BibitemShut
  {NoStop}%
\end{thebibliography}%
\end{document}